\documentclass[a4paper,11pt]{article}
\usepackage{jcappub}
\usepackage{hyperref}
\usepackage[T1]{fontenc}
\usepackage{amsthm}
\usepackage{scrextend}
\newtheorem*{theorem}{No-Go Theorem}
\usepackage{multicol}
\definecolor{darkblue}{rgb}{0.0, 0.0, 0.6}

\def\be{\begin{equation}}
\def\ee{\end{equation}}
\def\ba{\begin{eqnarray}}
\def\ea{\end{eqnarray}}
\renewcommand{\(}{\left(}
\renewcommand{\)}{\right)}
\renewcommand{\[}{\left[}
\renewcommand{\]}{\right]}
\newcommand{\p}{\partial}

\usepackage{color}

\newcommand{\fnl}{f_{\rm NL}}

\title{Matter bounce cosmology with a generalized single field: non-Gaussianity and an extended no-go theorem}

\author[a]{Yu-Bin Li,}
\author[b]{Jerome Quintin\footnote{Vanier Canada Graduate Scholar.},}
\author[c,d,a]{Dong-Gang Wang}
\author[a]{and Yi-Fu Cai}

\affiliation[a]{CAS Key Laboratory for Researches in Galaxies and Cosmology, Department of Astronomy, University of Science and Technology of China, Chinese Academy of Sciences,\\
Hefei, Anhui 230026, China}
\affiliation[b]{Department of Physics, McGill University,\\
3600 rue University, Montr\'eal, QC, H3A 2T8, Canada}
\affiliation[c]{Leiden Observatory, Leiden University,\\
2300 RA Leiden, The Netherlands}
\affiliation[d]{Lorentz Institute for Theoretical Physics, Leiden University,\\
2333 CA Leiden, The Netherlands}

\emailAdd{lyb2166@mail.ustc.edu.cn}
\emailAdd{jquintin@physics.mcgill.ca}
\emailAdd{wdgang@strw.leidenuniv.nl}
\emailAdd{yifucai@ustc.edu.cn}

\abstract{We extend the matter bounce scenario to a more general theory in which the background dynamics and cosmological perturbations are generated by a $k$-essence scalar field with an arbitrary sound speed. When the sound speed is small, the curvature perturbation is enhanced, and the tensor-to-scalar ratio, which is excessively large in the original model, can be sufficiently suppressed to be consistent with observational bounds. Then, we study the primordial three-point correlation function generated during the matter-dominated contraction stage and find that it only depends on the sound speed parameter. Similar to the canonical case, the shape of the bispectrum is mainly dominated by a local form, though for some specific sound speed values a new shape emerges and the scaling behaviour changes. Meanwhile, a small sound speed also results in a large amplitude of non-Gaussianities, which is disfavored by current observations. As a result, it does not seem possible to suppress the tensor-to-scalar ratio without amplifying the production of non-Gaussianities beyond current observational constraints (and vice versa). This suggests an extension of the previously conjectured no-go theorem in single field nonsingular matter bounce cosmologies, which rules out a large class of models. However, the non-Gaussianity results remain as a distinguishable signature of matter bounce cosmology and have the potential to be detected by observations in the near future.}

\begin{document}
\maketitle
\flushbottom

\section{Introduction}\label{sec:intro}

Matter bounce cosmology\ \cite{Brandenberger:2012zb}
is a very early universe structure formation scenario alternative to
the paradigm of inflationary cosmology (see, e.g.,\ \cite{Brandenberger:2011gk} for a review of inflation, its problems
and its alternatives).
The idea is that quantum fluctuations exit the Hubble radius in a matter-dominated contracting phase before the Big Bang,
which generates a scale-invariant power spectrum of curvature perturbations\ \cite{Wands:1998yp,Finelli:2001sr}.
The contracting phase is then followed by a bounce and the standard phases of hot Big Bang cosmology.
This construction solves the usual problems of standard Big Bang cosmology such as the horizon and flatness problems,
but in addition, it is free of the trans-Planckian corrections that plague inflationary cosmology\ \cite{Martin:2000xs}, and
one can naturally avoid reaching a singularity at the time of the Big Bang
(contrary to standard\footnote{The singularity before inflation could be avoided with, for example, bounce inflation
(e.g.,\ \cite{Wan:2015hya}).} inflation\ \cite{Borde:1993xh,Borde:2001nh})
under the assumption that new physics appears
at high energy scales\ \cite{Brandenberger:2011gk,Brandenberger:2012zb}.
Nonsingular bounces can be constructed in various ways using matter violating the Null Energy Condition (NEC),
with a modified gravity action, or within a quantum theory of gravity
(see the reviews\ \cite{Brandenberger:2012zb,Novello:2008ra,Cai:2014bea,Battefeld:2014uga,Brandenberger:2016vhg} and references therein).

A typical way of constructing a nonsingular matter bounce cosmology is to assume the existence of a new scalar field.
With a canonical Lagrangian, the oscillation of the scalar field can drive a matter-dominated contracting phase
when the ratio of the pressure to the energy density averages zero.
As the energy scale of the universe increases, new terms can appear in the Lagrangian that violate the NEC
and drive a nonsingular bounce. For example,
using a Galileon scalar field\ \cite{Nicolis:2008in} (or equivalently, in Horndeski theory\ \cite{Horndeski:1974wa}),
one can construct a stable NEC violating nonsingular bounce\ \cite{Qiu:2011cy,Easson:2011zy,Cai:2012va,Cai:2013vm,Osipov:2013ssa,Battarra:2014tga}
that may be free of ghost and gradient instabilities\ \cite{Ijjas:2016tpn,Ijjas:2016vtq}
(see, however, the difficulties in doing so as pointed out by\ \cite{Libanov:2016kfc,Kobayashi:2016xpl,Cai:2016thi,Creminelli:2016zwa}).

To distinguish the matter bounce scenario from inflation observationally, studying primordial non-Gaussianities is a useful
tool\footnote{Another observable quantity, besides non-Gaussianities, that would allow one to differentiate
between inflation and the matter bounce scenario
is the running of the scalar spectral index (see\ \cite{Lehners:2015mra,Cai:2016hea}).}.
In the case of inflation, after the calculation of the bispectra generated in single field slow-roll models\ \cite{Maldacena:2002vr},
there have been many studies in the past decade trying to extend the simplest result, which largely enriched the phenomenology of nonlinear
perturbations (see \cite{Chen:2010xka, Wang:2013eqj} for reviews).
In particular, one important progress has been to generalize the canonical inflaton to a $k$-essence scalar field
\cite{ArmendarizPicon:2000dh,ArmendarizPicon:2000ah}, such as $k$-inflation\ \cite{ArmendarizPicon:1999rj,Garriga:1999vw} and
DBI models\ \cite{Silverstein:2003hf,Alishahiha:2004eh}, which are collectively known as general single field inflation\ \cite{Chen:2006nt}.
In these models, due to the effects of a small sound speed, the amplitude of the bispectrum is enhanced and interesting shapes
emerge\ \cite{Seery:2005wm,Chen:2006nt,Cheung:2007st,Chen:2010xka,Noller:2011hd,Wang:2013eqj}.
In a matter-dominated contracting phase, the calculation of the bispectrum has only been done by\ \cite{Cai:2009fn} for the original
matter bounce model with a canonical scalar field.
A natural extension is thus to consider a $k$-essence scalar field\footnote{This could be easily
further generalized to a Galileon field\ \cite{Deffayet:2011gz}, which has also been done for inflation
(see, e.g.,\ \cite{Kobayashi:2010cm,Burrage:2010cu,Creminelli:2010qf,Kobayashi:2011nu,Gao:2011qe}).}
similarly to what has been done in inflationary cosmology,
especially since the appearance of a noncanonical field is quite common in the literature of nonsingular bouncing cosmology
in order to violate the NEC as explained above.
Because the perturbations behave differently in matter bounce cosmology compared to inflation,
in particular due to the growth of curvature perturbations on super-Hubble scales during the matter-dominated contracting phase,
the canonical matter bounce yields non-Gaussianities with negative sign and order one amplitude,
which differs from the results in canonical single field inflation.
It would be interesting to explore how these non-Gaussianity results change
when one generalizes the original matter bounce scenario to be based on a $k$-essence scalar field.

Besides non-Gaussianity, another interesting observable for very early universe models is the tensor-to-scalar ratio $r$.
In the original matter bounce scenario, this ratio is predicted to be very large\ \cite{Cai:2008qw,Cai:2014xxa}.
Indeed, the scalar and tensor power spectra share the same amplitude, and accordingly,
the tensor-to-scalar ratio is naturally of order unity\ \cite{Quintin:2015rta}.
This is well beyond the current observational bound from the Cosmic Microwave Background (CMB),
which states that $r<0.07$ at $95\%$ confidence\ \cite{Array:2015xqh}.

A resolution to this problem is to allow for the growth of curvature perturbations during the bounce phase,
which suppresses the tensor-to-scalar ratio. However, curvature perturbations tend to remain constant through
the bounce phase on super-Hubble scales\ \cite{Xue:2013bva,Battarra:2014tga}. In fact, amplification can only be achieved under
some tuning of the parameters, and the overall growth is still
limited\footnote{The studies of Refs.\ \cite{Xue:2013bva,Battarra:2014tga,Quintin:2015rta}
have been carried out for models where the nonsingular bounce is attributed to a noncanonical scalar field.
Loop quantum cosmology (LQC) provides an alternative class of nonsingular bouncing models that could suppress $r$ during the bounce.
In LQC, the amplitude of the suppression depends on the equation of state during the bounce;
if it is close to zero, then the suppression is very strong (see\ \cite{Cai:2014xxa,Cai:2014jla,Wilson-Ewing:2015sfx}
and references therein for a discussion of LQC effects in nonsingular bouncing cosmology).}\ \cite{Quintin:2015rta}.
Yet, if the scalar modes are amplified, another problem follows in that it leads to the production of large
non-Gaussianities\ \cite{Quintin:2015rta}, a problem that might be generic to a large class of nonsingular bounces\ \cite{Gao:2014hea,Gao:2014eaa}.
Again, these large non-Gaussianities are excluded by current measurements from the CMB\ \cite{Ade:2015ava}.
This leads to conjecture that single field matter bounce cosmology suffers from a no-go theorem\ \cite{Quintin:2015rta},
which states that one cannot satisfy the bound on $r$ without violating the bounds on non-Gaussianities and vice versa.

There is another way to suppress the tensor-to-scalar ratio if the sound speed of the perturbations can be smaller than the speed of light
during the matter-dominated contracting phase.
For example, in the $\Lambda$CDM bounce scenario\ \cite{Cai:2014jla} (and its extension\ \cite{Cai:2015vzv}; see the review\ \cite{Cai:2016hea}),
if there exists a form of dark matter with a small sound speed that dominates the contracting phase
when the scale-invariant power spectra are generated, then the tensor-to-scalar ratio is already suppressed proportionally to the sound speed.
Therefore, this provides another motivation to study non-Gaussianities when the sound speed is small during the matter-dominated contracting phase.
An immediate question is whether the no-go theorem still holds true in this case or whether it can be circumvented.
In this work, we want to explore this possibility of having a $k$-essence scalar field
that would mimic dust-like matter with a small sound speed at low energies and that could play
the role of the NEC violating scalar field during the bounce.

In this paper, we will evaluate the bispectrum produced by a $k$-essence scalar field in a matter-dominated contracting universe.
This more general setup will yield richer features, which have the potential to be detected by future non-Gaussianity observations.
In particular, the shapes, amplitudes, and scaling behaviors will be studied systematically.
We will show that a small sound speed implies a large amplitude associated with the three-point function.
Accordingly, we will claim that the no-go theorem is not circumvented but rather extended:
in single field matter bounce cosmology, one cannot suppress the tensor-to-scalar ratio, either from the onset of the initial conditions
in the matter contracting phase or from the amplification of the curvature perturbations during the bouncing phase,
without producing large non-Gaussianities.

The outline of the paper is as follows.
In section\ \ref{sec:setup}, we first introduce the background dynamics of the matter bounce scenario
and introduce the class of $k$-essence scalar field models that we study in this paper.
In section\ \ref{sec:2ptfunction}, we calculate the power spectra of curvature perturbations and tensor modes
and show how a small sound speed
coming from the $k$-essence scalar field allows for the suppression of the tensor-to-scalar ratio.
We then consider the primordial non-Gaussianity in section\ \ref{sec:NG}.
Using the in-in formalism, we evaluate every contribution to the three-point function
and give a detailed analysis of the size and shapes of the resulting bispectrum.
In section\ \ref{sec:fNLandNoGo}, we compute the amplitude parameter of non-Gaussianities in different limits
and finally combine these results with the bound on the sound speed from section\ \ref{sec:2ptfunction}
to show that the no-go theorem in matter bounce cosmology is extended.
We summarize our results in section\ \ref{sec:conclusion}.
Throughout this paper, we use the mostly plus metric convention, and we define the reduced Planck mass
to be $M_\mathrm{Pl}=(8\pi G_\mathrm{N})^{-1/2}$, where $G_\mathrm{N}$ is Newton's gravitational constant.

\section{Setup and background dynamics}\label{sec:setup}

The idea of the matter bounce scenario is to begin with a matter-dominated contracting phase.
At the background level, this corresponds to having a scale factor as a function of physical time given by
\begin{equation}
 a(t)=a_B\left(\frac{t-\tilde{t}_B}{t_B-\tilde{t}_B}\right)^{2/3}~,
\end{equation}
and the Hubble parameter follows,
\be
H(t)=\frac{2}{3(t-\tilde t_B)}~,
\ee
where $t_B$ corresponds to the time of the beginning of the bounce phase and $\tilde t_B$ corresponds to the time at which the singularity
would occur if no new physics appeared at high energy scales. Accordingly, $a_B$ is the value of the scale factor at $t_B$.
In terms of the conformal time $\tau$ defined by $\mathrm{d}\tau=a^{-1}\mathrm{d}t$, the scale factor is given by
\begin{equation}
 a(\tau)=a_B\left(\frac{\tau-\tilde\tau_B}{\tau_B-\tilde{\tau}_B}\right)^2~,
\end{equation}
where $\tau_B$ and $\tilde\tau_B$ are the conformal times corresponding to $t_B$ and $\tilde t_B$.
Throughout the rest of this paper, the scale factor is normalized such that $a_B=1$.

One can define the usual ``slow-roll'' parameters of inflation by
\begin{equation}
 \epsilon\equiv-\frac{\dot H}{H^2}=\frac{3}{2}(1+w)~,\qquad\eta\equiv\frac{\dot\epsilon}{H\epsilon}~,
\end{equation}
where a dot denotes a derivative with respect to physical time, and $w\equiv p/\rho$ is the equation of state parameter
with $p$ and $\rho$ denoting pressure and energy density, respectively.
In the case of the matter bounce, the matter contracting phase implies that pressure vanishes,
which is to say that
\begin{equation}
\label{eq:wepsetamatter}
 w=0~,\qquad\epsilon=\frac{3}{2}~,\qquad\eta=0~.
\end{equation}
If the pressure does not vanish exactly but is still very small, i.e.~$|p/\rho|\ll 1$, then
the values for $w$, $\epsilon$, and $\eta$ in equation\ \eqref{eq:wepsetamatter} are only valid as leading order approximations,
and they will be time dependent rather than constant. In this paper, we will work in the limit where equation\ \eqref{eq:wepsetamatter}
is valid.

In the usual matter bounce scenario, one would introduce a canonical scalar
field to drive the matter-dominated contracting phase and describe the cosmological fluctuations.
In this paper, we aim for more generality and assume that
the perturbations are introduced by a $k$-essence scale field $\phi$ with Lagrangian density
of the form\footnote{For an introduction to such a Lagrangian in early universe cosmology
with the derivation of the background equations of motion and the definition of the different parameters,
see, e.g.,\ \cite{ArmendarizPicon:1999rj,Garriga:1999vw,Seery:2005wm,Chen:2006nt}.}
\begin{equation}
\label{eq:Lagrangian}
 \mathcal{L}_\phi=P(X,\phi)~,
\end{equation}
where $X\equiv-\partial_\mu\phi\partial^\mu\phi/2$,
and we assume that the scalar field is minimally coupled to gravity.
The energy density and pressure of this scalar field are then given by
\be
\label{eq:defrhop}
 \rho=2X P_{,X}-P~,\qquad p=P~,
\ee
where a comma denotes a partial derivative, e.g.~$P_{,X}\equiv\partial P/\partial X$.
Thus, the Friedmann equations read
\be
 3M_\mathrm{Pl}^2H^2=2X P_{,X}-P~,\qquad M_\mathrm{Pl}^2\dot H=-X P_{,X}~.
\ee
Since we want a matter-dominated contracting phase, the pressure of the  scalar field should vanish (at least in average),
and $\rho=2X P_{,X}\propto a^{-3}$.

It is helpful to have one specific example where a $k$-essence field drives the matter contraction. Let us consider the following Lagrangian density:
\be \label{example}
\mathcal{L}_\phi=K(X)=\frac{1}{8}(X-c^2)^2~.
\ee
This type of Lagrangian belongs to a subclass of $k$-essence models $P(X,\phi)$ where the kinetic terms $K(X)$ are separate
from the potential terms $V(\phi)$, i.e.~$P(X,\phi)=K(X)-V(\phi)$. Moreover, the above Lagrangian has vanishing potential.
Then, the ghost condensate solution is given by $X=c^2$ and $\phi(t)=ct+\pi(t)$, with $\dot\pi(t)\ll c$. In this case,
the background equations yield $p\simeq 0$ and $\rho\sim\dot\pi\propto a^{-3}$, which exactly corresponds to a matter-dominated universe.
More details about this model can be found in \cite{Lin:2010pf}.
We note that there should be also other forms of $P(X,\phi)$ that can drive a matter contraction,
and remarkably, the analysis that follows in this paper is done in a
{\it model-independent} way and does {\it not} rely on the specific model of equation\ \eqref{example}.

The sound speed and another ``slow-roll'' parameter are defined by\footnote{We assume that the cosmological perturbations
will remain adiabatic throughout the matter-dominated contracting phase.}
\begin{equation}
\label{eq:defcs2s}
 c_\mathrm{s}^2\equiv\frac{\partial p}{\partial\rho}=\frac{P_{,X}}{P_{,X}+2XP_{,XX}}~,\qquad s\equiv\frac{\dot c_\mathrm{s}}{c_\mathrm{s}H}~.
\end{equation}
Calculations will be done for a general sound speed, but as we will argue, we will be interested in the small sound speed limit,
which can be realized with the appropriate form for $P(X,\phi)$. For instance, the explicit example given by
equation\ \eqref{example} yields $c_\mathrm{s}\simeq\dot\pi/c\ll1$.
Furthermore, we will generally assume later that the sound speed
remains nearly constant, which is to say that $|s|\ll 1$.
We also define two other variables for later convenience,
\begin{equation}
\label{eq:Sigmadef}
 \Sigma\equiv XP_{,X}+2X^2P_{,XX}=\frac{M_\mathrm{Pl}^2H^2\epsilon}{c_\mathrm{s}^2}~,
\end{equation}
and
\begin{equation}
\label{eq:lambdadef}
 \lambda\equiv X^2P_{,XX}+\frac23 X^3P_{,XXX}=\frac{X}{3} \Sigma_{,X}-\frac13\Sigma~.
\end{equation}
The ratio $\lambda/\Sigma$ will be of particular interest in the following sections.
For inflation, it depends on the specific realization of the general single field, such as DBI and $k$-inflation models.
For the matter bounce scenario, it can be obtained in an approximately model-independent way.
The detailed calculation is in Appendix\ \ref{sec:ratiolambdasigma}, where we find that the ratio
$\lambda/\Sigma$ can be expressed in terms of the sound speed, as shown by equation\ \eqref{eq:lambdaoSigmaresult}.

\section{Mode functions and two-point correlation functions}\label{sec:2ptfunction}

We begin with an action of the form
\begin{equation}
\label{eq:action0}
 S=\int\mathrm{d}^4x~\sqrt{-g}\left(\frac{1}{2}M_\mathrm{Pl}^2R+\mathcal{L}_\phi\right)~,
\end{equation}
where $g$ is the determinant of the metric tensor and $R$ is the Ricci scalar.
Importantly, we assume that the matter Lagrangian $\mathcal{L}_\phi$ has the general form of equation\ \eqref{eq:Lagrangian},
but we do not restrict our attention to any specific model.
By perturbing up to second order the above action,
one finds\footnote{Again, see, e.g.,\ \cite{Garriga:1999vw,Seery:2005wm,Chen:2006nt,Chen:2010xka}
for a derivation of the perturbation equations in $k$-essence early universe cosmology.}
\begin{equation}
 S^{(2)}=\int\mathrm{d}\tau\mathrm{d}^3\vec{x}~z^2\left[\zeta'^2-c_\mathrm{s}^2(\vec{\nabla}\zeta)^2\right]~,
\end{equation}
where $\zeta(\tau,\vec{x})$ denotes the curvature perturbation in the comoving gauge, i.e.~on
slices where fluctuations of the scalar field vanish ($\delta\phi=0$).
Also, a prime represents a derivative with respect to conformal time,
$\vec{\nabla}=\partial_i$ is the spatial gradient, and we define $z^2\equiv 2\epsilon a^2M_\mathrm{Pl}^2/c_\mathrm{s}^2$.
Transforming to Fourier space, the second-order perturbed action becomes
\begin{equation}
 S^{(2)}=\int\mathrm{d}\tau\int\frac{\mathrm{d}^3\vec{k}}{(2\pi)^3}~z^2
 \left[\zeta'(\vec{k})\zeta'(-\vec{k})-c_\mathrm{s}^2k^2\zeta(\vec{k})\zeta(-\vec{k})\right]~,
\end{equation}
where $k^2\equiv\vec{k}\cdot\vec{k}=|\vec{k}|^2$.
Upon quantization of the curvature perturbation, one has
\begin{equation}
 \hat\zeta(\tau,\vec{k})=\hat a^{\dagger}_{\vec{k}}u_k(\tau)+\hat a_{-\vec{k}}u^{\ast}_k(\tau)~,
\end{equation}
where the annihilation and creation operators satisfy
the usual commutation relation
$[\hat a_{\vec{k}},\hat a^{\dagger}_{\vec{k}'}]=(2\pi)^3\delta^{(3)}(\vec{k}-\vec{k}')$.
The equation of motion of the mode function is then given by
\begin{equation}
\label{eq:vkEoM}
 v_{k}''+\left(c_\mathrm{s}^2k^2-\frac{z''}{z}\right)v_{k}=0~,
\end{equation}
where the mode function is rescaled as $v_k=zu_k$
($v_k$ is called the Mukhanov-Sasaki variable).
Together with the commutation relation $[\hat\zeta(\vec{k}_1),\hat\zeta'(\vec{k}_2)]=(2\pi)^3\delta^{(3)}(\vec{k}_1+\vec{k}_2)$,
one finds (see, e.g.,\ \cite{Cai:2009fn})
\begin{align}
\label{eq:uksol}
 u_k(\tau)&=\frac{iA[1-ic_\mathrm{s}k(\tau-\tilde{\tau}_B)]}{2\sqrt{\epsilon c_\mathrm{s}k^3}(\tau-\tilde{\tau}_B)^3}
  e^{ic_\mathrm{s}k(\tau-\tilde{\tau}_B)} \\
\label{eq:ukprimesol}
 u'_k(\tau)&=\frac{iA}{2\sqrt{\epsilon c_\mathrm{s}k^3}}
  \left(\frac{-3[1-ic_\mathrm{s}k(\tau-\tilde{\tau}_B)]}{(\tau-\tilde{\tau}_B)^4}
  +\frac{c_\mathrm{s}^2k^2}{(\tau-\tilde\tau_B)^2}\right)
  e^{ic_\mathrm{s}k(\tau-\tilde{\tau}_B)}
\end{align}
to be the solution to the equation of motion\ \eqref{eq:vkEoM} in the context of a matter-dominated contracting universe
as described in the previous section. Here, $A$ is a normalization constant that is determined
by the quantum vacuum condition at Hubble radius crossing in the contracting phase,
which is given by $A=(\tau_B-\tilde\tau_B)^2/M_\mathrm{Pl}$.

The general two-point correlation functions are given by
\begin{align}
 \langle\hat\zeta(\tau_1,\vec{k}_1)\hat\zeta(\tau_2,\vec{k}_2)\rangle
  &=(2\pi)^3\delta(\vec{k}_1+\vec{k}_2)u^\ast_{k_1}(\tau_1)u_{k_1}(\tau_2)~, \\
 \langle\hat\zeta(\tau_1,\vec{k}_1)\hat\zeta'(\tau_2,\vec{k}_2)\rangle
  &=(2\pi)^3\delta(\vec{k}_1+\vec{k}_2)u^{\ast}_{k_1}(\tau_1)u'_{k_1}(\tau_2)~,
\end{align}
and in particular, the power spectrum, evaluated at the bounce point $\tau_B$ (well after Hubble radius exit), is given by
\begin{equation}
 \langle\hat\zeta(\tau_B,\vec{k})\hat\zeta(\tau_B,\vec{k}')\rangle=(2\pi)^3\delta^{(3)}(\vec{k}+\vec{k}')\frac{2\pi^2}{k^3}
 \mathcal{P}_{\zeta}(\tau_B,k)~,
\end{equation}
where
\begin{equation}
\label{eq:Pzeta}
 \mathcal{P}_{\zeta}(\tau_B,k)=\frac{A^2}{8\pi^2\epsilon c_\mathrm{s}(\tau_B-\tilde\tau_B)^6}
 =\frac{1}{12\pi^2c_\mathrm{s}M_\mathrm{Pl}^2(\tau_B-\tilde\tau_B)^2}~.
\end{equation}
The scale invariance of the power spectrum in matter bounce cosmology is thus explicit from the above.

The above focused only on the scalar perturbations, but as mentioned in the introduction, the matter bounce scenario
also generates a scale-invariant power spectrum of tensor perturbations. Considering the transverse and
traceless perturbations to the spatial metric, $\delta g_{ij}=a^2 h_{ij}$,
which can be decomposed as
\begin{equation}
 h_{ij}(\tau,\vec{x})=h_+(\tau,\vec{x})e_{ij}^++h_\times(\tau,\vec{x})e_{ij}^\times
\end{equation}
with two fixed polarization tensors $e_{ij}^+$ and $e_{ij}^\times$,
the second-order perturbed action has contributions of the form
\begin{equation}
 S^{(2)}\supset\frac{M_\mathrm{Pl}^2}{4}\int\mathrm{d}\tau\mathrm{d}^3\vec{x}~a^2\left[h'^2-(\vec{\nabla}h)^2\right]
\end{equation}
for each polarization state $h_+$ and $h_\times$. By normalizing each state as $\mu=aM_\mathrm{Pl}h/2$,
the second-order perturbed action is of canonical form ($\mu$ is the Mukhanov-Sasaki variable),
and the resulting equation of motion for each state is
\begin{equation}
 \mu_k''+\left(k^2-\frac{a''}{a}\right)\mu_k=0~,
\end{equation}
where the equation is written in Fourier space.
Since $a\sim\tau^2$ in a matter-dominated contracting phase, one has $a''/a=2/\tau^2$, and so, one expects a scale-invariant power spectrum
just as in de Sitter space.
The tensor power spectrum is given by
\begin{equation}
 \mathcal{P}_\mathrm{t}=2\mathcal{P}_h=2\left(\frac{2}{aM_\mathrm{Pl}}\right)^2\frac{k^3}{2\pi^2}|\mu_k|^2~,
\end{equation}
where the first factor of 2 accounts for the two polarizations $+$ and $\times$,
and the factor $[2/(aM_\mathrm{Pl})]^2$ comes from the normalization of $\mu$.
Upon matching with quantum vacuum initial conditions
at Hubble radius crossing similar to the above treatment for scalar modes,
one finds the power spectrum of tensor modes at the bounce point to be given by
\begin{equation}
\label{eq:Pt}
 \mathcal{P}_\mathrm{t}(\tau_B,k)=\frac{2}{\pi^2M_\mathrm{Pl}^2(\tau_B-\tilde\tau_B)^2}~,
\end{equation}
which is indeed independent of scale.

The tensor-to-scalar ratio is then defined to be
\begin{equation}
 r\equiv\frac{\mathcal{P}_\mathrm{t}}{\mathcal{P}_\zeta}~.
\end{equation}
It follows from equations\ \eqref{eq:Pzeta} and\ \eqref{eq:Pt} that
\begin{equation}
 r=24c_\mathrm{s}
\end{equation}
in the context of matter bounce cosmology with a general $k$-essence scalar field\footnote{Of course,
this assumes that the perturbations remain constant on super-Hubble scales after the matter contraction phase,
in particular through the bounce and until the beginning of the radiation-dominated expanding phase of standard
Big Bang cosmology.}.
On one hand, this highlights the problem of standard matter bounce cosmology, which is driven by a canonical scalar field with $c_\mathrm{s}=1$,
in which case $r=24$.
On the other hand, the above result provides a natural mechanism to suppress the tensor-to-scalar ratio
provided the $k$-essence scalar field has an appropriately small sound speed.
For example, satisfying the observational bound\ \cite{Array:2015xqh} $r<0.07$ at $95\%$ confidence imposes a bound
on the sound speed of the order of
\begin{equation}
 c_\mathrm{s}\lesssim 0.0029~.
\end{equation}

\section{Non-Gaussianity}\label{sec:NG}

The previous section showed that a $k$-essence scalar field could yield a small tensor-to-scalar ratio in the context
of the matter bounce scenario. This is done at the expense of having a small sound speed. In what follows, the goal
is to compute the bispectrum and see how a small sound speed affects the results.

\subsection{Cubic action}\label{sec:S3}

To evaluate the three-point correlation function, we must expand the action\ \eqref{eq:action0}
up to third order.
Let us recall the result of\ \cite{Chen:2006nt},
the third-order interaction action of a general single scalar field\footnote{From here on, we take $M_\mathrm{Pl}=1$ for convenience.},
\begin{align}
\label{action3}
 S^{(3)}=&\int\mathrm{d}t\mathrm{d}^3\vec{x}~\Big\{
  -a^3\Big[\Sigma\Big(1-\frac{1}{c_\mathrm{s}^2}\Big)+2\lambda\Big]\frac{\dot{\zeta}^3}{H^3}
  +\frac{a^3\epsilon}{c_\mathrm{s}^4}(\epsilon-3+3c_\mathrm{s}^2)\zeta\dot{\zeta}^2 \nonumber \\
  &+\frac{a\epsilon}{c_\mathrm{s}^2}(\epsilon-2s+1-c_\mathrm{s}^2)\zeta(\partial\zeta)^2
  -2a\frac{\epsilon}{c_\mathrm{s}^2}\dot{\zeta}(\partial\zeta)(\partial \chi)+\frac{a^3\epsilon}{2c_\mathrm{s}^2}\frac{\mathrm{d}}{\mathrm{d}t}\left(\frac{\eta}{c_\mathrm{s}^2}\right)\zeta^2\dot{\zeta} \nonumber \\
  & +\frac{\epsilon}{2a}(\partial\zeta)(\partial\chi)\partial^2\chi
 +\frac{\epsilon}{4a}(\partial^2\zeta)(\partial\chi)^2+2f(\zeta)\left.\frac{\delta L}{\delta \zeta}\right|_1\Big\}~,
\end{align}
where it is understood that $(\partial\zeta)^2=\partial_i\zeta\partial^i\zeta$,
$(\partial\zeta)(\partial\chi)=\partial_i\zeta\partial^i\chi$, $\partial^2\zeta=\partial_i\partial^i\zeta$,
and where we define $\chi$ such that $\p^2\chi=a^2\epsilon\dot\zeta$. Also, we have
\begin{equation}
 \left.\frac{\delta L}{\delta\zeta}\right|_1=a\left(\frac{\mathrm{d}\partial^2\chi}{\mathrm{d}t}+H\partial^2\chi-\epsilon\partial^2\zeta\right)~,
\end{equation}
\begin{align}
\label{redefinition}
 f(\zeta)=&~\frac{\eta}{4c_\mathrm{s}^2}\zeta^2+\frac{1}{c_\mathrm{s}^2H}\zeta\dot{\zeta}
  +\frac{1}{4a^2H^2}\{-(\partial\zeta)(\partial\zeta)+\partial^{-2}[\partial_i\partial_j(\partial^i\zeta\partial^j\zeta)]\} \nonumber \\
   &+\frac{1}{2a^2H}\{(\partial\zeta)(\partial\chi)-\partial^{-2}[\partial_i\partial_j(\partial^i\zeta\partial^j\chi)]\}~,
\end{align}
where $\partial^{-2}$ is the inverse Laplacian.

The first and second terms in the last line of equation\ \eqref{action3} can be reexpressed as
\be
 \frac{\epsilon}{2a}(\partial\zeta)(\partial\chi) \partial^2 \chi +\frac{\epsilon}{4a}(\partial^2\zeta)(\partial\chi)^2=
 -\frac{a^3\epsilon^3}{2}\zeta\dot\zeta^2+\frac{\epsilon}{2a}\zeta(\p_i\p_j\chi)(\p^i\p^j\chi)+\mathcal{K}~,
\ee
where the boundary term is given by
\be
 \mathcal{K}=\p_i\[\zeta(\p^i\chi)(\p^2\chi)+\frac{1}{2}(\p^i\zeta)(\p\chi)^2-\zeta(\p^i\p^j\chi)(\p_j\chi)\]~.
\ee
Since the $\p_i[...]$ term above does not contribute to the three-point function, the third-order action, equation\ \eqref{action3},
is equivalent to
\begin{align}
\label{eq:S3int}
 S^{(3)}=&\int\mathrm{d}t\mathrm{d}^3\vec{x}~\Big\{
  -a^3\Big[\Sigma\Big(1-\frac{1}{c_\mathrm{s}^2}\Big)+2\lambda\Big]\frac{\dot{\zeta}^3}{H^3}
  +\frac{a^3\epsilon}{c_\mathrm{s}^4}(\epsilon-3+3c_\mathrm{s}^2)\zeta\dot{\zeta}^2 \nonumber \\
  &+\frac{a\epsilon}{c_\mathrm{s}^2}(\epsilon-2s+1-c_\mathrm{s}^2)\zeta(\partial\zeta)^2
  -2a\frac{\epsilon}{c_\mathrm{s}^2}\dot{\zeta}(\partial\zeta)(\partial \chi) +\frac{a^3\epsilon}{2c_\mathrm{s}^2}\frac{\mathrm{d}}{\mathrm{d}t}\left(\frac{\eta}{c_\mathrm{s}^2}\right)\zeta^2\dot{\zeta}\nonumber \\
  &  -\frac{a^3\epsilon^3}{2}\zeta\dot\zeta^2
  +\frac{\epsilon}{2a}\zeta(\p_i\p_j\chi)(\p^i\p^j\chi)+ 2 f(\zeta)\left.\frac{\delta L}{\delta \zeta}\right|_1\Big\}~.
\end{align}
In the case of a canonical field with $c_\mathrm{s}=1$, this action returns to equation (15) of\ \cite{Cai:2009fn}. Meanwhile, as usual the last term in this action is removed by performing the field redefinition
\be
 \zeta\rightarrow\tilde\zeta+f(\tilde\zeta)~,
\ee
where $\tilde\zeta$ denotes the field after redefinition.

\subsection{Contributions to the shape function}\label{sec:A}

In this section, we calculate the three-point correlation function using the in-in formalism (to leading order in perturbation theory;
see, e.g.,\ \cite{Maldacena:2002vr,Chen:2010xka,Wang:2013eqj} for the methodology),
\begin{equation} \label{inin}
 \langle O(t)\rangle=-2~\mathrm{Im}\int^t_{-\infty}\mathrm{d}\bar t~\langle0|O(t)L_{\rm int}(\bar{t})|0\rangle~,
\end{equation}
where $O$ represents a set of operators of the form $\hat\zeta^3$ in our case of interest.
Then, the shape function, $\mathcal{A}$, is defined such that\footnote{We use $\zeta_{\vec{k}_i}$
to refer to $\hat\zeta(\tau,\vec{k}_i)$ to simplify the notation from here on.}
\begin{equation}
 \langle\zeta_{\vec{k}_1}\zeta_{\vec{k}_2}\zeta_{\vec{k}_3}\rangle
 =(2\pi)^7\delta^{(3)}\Big(\sum_i\vec{k}_i\Big)\frac{\mathcal{P}_{\zeta}^2}{\prod_i k_i^3}\mathcal{A}(\vec{k}_1,\vec{k}_2,\vec{k}_3)~.
\end{equation}
In what follows, we list all the contributions to the shape function coming from the field redefinition
and the interaction action\ \eqref{eq:S3int}. It is easy to check that, when taking the limit $c_\mathrm{s}=1$,
one recovers the results of\ \cite{Cai:2009fn} for the matter bounce with a canonical scalar field as expected.

\subsubsection{Contribution from the field redefinition}

  In momentum space, the field redefinition can be written as
  \begin{gather}
  \zeta_{\vec k}\rightarrow\tilde\zeta_{\vec k}+\int\frac{\mathrm{d}^3\vec{k}_1}{(2\pi)^3}\left[-\frac{3}{2c_\mathrm{s}^2}
   -\frac{3\epsilon}{4}\left(\frac{\vec{k}_1\cdot(\vec k-\vec{k}_1)}{k_1^2}
   -\frac{(\vec k\cdot\vec{k}_1)[\vec k\cdot(\vec k-\vec{k}_1)]}{k^2k_1^2}\right)\right]\tilde\zeta_{\vec k_1}\tilde\zeta_{\vec k-\vec k_1}~.
  \end{gather}
  This redefinition has the following contribution to the three-point correlation function,
  \begin{align}
   \langle\zeta_{\vec{k}_1}\zeta_{\vec{k}_2}\zeta_{\vec{k}_3}\rangle_\mathrm{redef}
    =&\int\frac{\mathrm{d}^3\vec{k}'}{(2\pi)^3}\left[-\frac{3}{2c_\mathrm{s}^2}
     -\frac{3\epsilon}{4}\left(\frac{\vec{k}'\cdot(\vec k_3-\vec{k}')}{k'^2}
     -\frac{(\vec k_3\cdot\vec{k}')(\vec k_3\cdot[\vec k_3-\vec{k}')]}{k_3^2k'^2}\right)\right] \nonumber \\
    &\times\left(\zeta_{\vec{k}_1}\zeta_{\vec{k}_2}\zeta_{\vec{k}'}\zeta_{\vec{k}_3-\vec{k}'}\right)
     +(2~{\rm permutations})~,
  \end{align}
  and accordingly, the contribution to the shape function is
  \begin{equation}
   \mathcal A_\mathrm{redef}=\Big(\frac{3\epsilon}{16}-\frac{3}{4c_\mathrm{s}^2}\Big)\sum_i k_i^3+\frac{3\epsilon}{64}\sum_{i\neq j}k_ik_j^2
   -\frac{3\epsilon}{64\prod_i k_i^2}\Big(\sum_{i\neq j}k_i^7k_j^2+\sum_{i\neq j}k_i^6k_j^3-2\sum_{i\neq j}k_i^5k_j^4\Big)~.
  \end{equation}
 When $c_\mathrm{s}^2\ll 1$,  this contribution is enhanced compared to the canonical case.

\subsubsection{Contribution from the $\zeta\dot{\zeta}^2$ term}

  The term $\zeta\dot{\zeta}^2$ in equation\ \eqref{eq:S3int} yields the following contribution to the bispectrum
  \begin{align}
   \langle\zeta_{\vec{k}_1}\zeta_{\vec{k}_2}\zeta_{\vec{k}_3}\rangle_{\zeta\dot{\zeta}^2}
   =&-2\times2~\mathrm{Im}\int_{-\infty}^{\tau_B}\mathrm{d}\bar\tau~(2\pi)^3\delta\Big(\sum_i\vec{k}_i\Big)
    a^2\Big[\frac{\epsilon}{c_\mathrm{s}^4}(\epsilon-3+3c_\mathrm{s}^2)-\frac{\epsilon^3}{2}\Big] \nonumber \\
   &\times u^\ast_{k_1}(\tau_B)u_{k_1}(\bar\tau)u^\ast_{k_2}(\tau_B)u'_{k_2}(\bar\tau)u^\ast_{k_3}(\tau_B)u'_{k_3}(\bar\tau)
   +(2~{\rm permutations}).
  \end{align}
  To leading order in $c_\mathrm{s}k_i(\tau_B-\tilde{\tau}_B)\ll 1$,
  i.e.~on scales larger than the sound Hubble radius\footnote{This is also called the
  Jeans radius; see\ \cite{Cai:2014jla,Quintin:2016qro} for an explicit definition of this scale and its role in matter
  bounce cosmology when $c_\mathrm{s}\neq 1$.},
  and recalling the solutions for $u_k$ and $u_k'$ [equations\ \eqref{eq:uksol} and\ \eqref{eq:ukprimesol}],
  we get the following contribution to the shape function,
  \begin{gather}
   \mathcal{A}_{\zeta\dot{\zeta}^2}=-\frac{c_\mathrm{s}^2}{8}
   \left[\frac{1}{c_\mathrm{s}^4}(\epsilon-3+3c_\mathrm{s}^2)-\frac{\epsilon^2}{2}\right]\sum_i k_i^3~.
  \end{gather}
  Again, when $c_\mathrm{s}^2\ll 1$, this contribution is enhanced compared to the canonical case.

\subsubsection{Contribution from the $\dot{\zeta}\partial\zeta\partial\chi$ term}

  A similar computation for this term yields the following contribution to the shape function
  \begin{align}
   \mathcal{A}_{\dot{\zeta}\partial\zeta\partial\chi}=-\frac{\epsilon}{8}\sum_i k_i^3+\frac{\epsilon}{8\prod_i k_i^2}\Big(\sum_{i\neq j}k_i^7k_j^2-\sum_{i\neq j}k_i^4k_j^5\Big).
  \end{align}
  We note that this contribution is independent of $c_\mathrm{s}$.

\subsubsection{Contribution from the $\zeta(\partial_i\partial_j\chi)^2$ term}

For this term, the contribution to the shape function is given by
  \begin{align}
   \mathcal{A}_{\zeta(\partial_i\partial_j\chi)^2}
   =&-\frac{c_\mathrm{s}^2\epsilon^2}{32}\sum_i k_i^3+\frac{c_\mathrm{s}^2\epsilon^2}{64}\sum_{i\neq j}k_i^2k_j \nonumber \\
   &+\frac{c_\mathrm{s}^2\epsilon^2}{64\prod_i k_i^2}\Big(\sum_i k_i^9-\sum_{i\neq j}k_i^6k_j^3+3\sum_{i\neq j}k_i^5k_j^4-3\sum_{i\neq j}k_i^7k_j^2\Big)~.
  \end{align}
  When $c_\mathrm{s}^2\ll 1$, this contribution is suppressed compared to the canonical case.

\subsubsection{Contribution from the $\dot\zeta^3$ term}

  The $\dot\zeta^3$ term is a new element in the Lagrangian caused by the nontrivial sound speed, which does not show up in the cubic action of canonical fields.
  Its contribution to the bispectrum is
  \begin{align}
   \langle\zeta_{\vec{k}_1}\zeta_{\vec{k}_2}\zeta_{\vec{k}_3}\rangle_{\dot\zeta^3}
    =&-6\times 2~\mathrm{Im}\int_{-\infty}^{\tau_B}\mathrm{d}\bar\tau~(2\pi)^3\delta^{(3)}\Big(\sum_i\vec{k}_i\Big)
     \Big(-\frac{aM_\mathrm{Pl}^2\epsilon}{Hc_\mathrm{s}^2}\Big)\Big(1-\frac{1}{c_\mathrm{s}^2}+2\frac{\lambda}{\Sigma}\Big) \nonumber \\
    &\times u^\ast_{k_1}(\tau_B)u'_{k_1}(\bar\tau)u^\ast_{k_2}(\tau_B)u'_{k_2}(\bar\tau)u^\ast_{k_3}(\tau_B)u'_{k_3}(\bar\tau)~,
  \end{align}
  where we have used the expression for $\Sigma$, equation \eqref{eq:Sigmadef}.
  Then the contribution to the shape function is expressed as
  \begin{equation}
   \mathcal{A}_{\dot\zeta^3}=-\frac{9}{2}\left(1-\frac{1}{c_\mathrm{s}^2}+2\frac{\lambda}{\Sigma}\right)\sum_ik_i^3~.
  \end{equation}
  Since this is a new contribution compared to the canonical case, it vanishes for $c_\mathrm{s}^2=1$.
  Indeed, when $c_\mathrm{s}^2=1$, $\lambda/\Sigma\simeq(1-c_\mathrm{s}^2)/(6c_\mathrm{s}^2)=0$
  (see equation\ \eqref{eq:lambdaoSigmaresult} in Appendix\ \ref{sec:ratiolambdasigma}) and $1-1/c_\mathrm{s}^2=0$.
  We note though that when $c_\mathrm{s}^2\ll 1$, this contribution is large.

\subsubsection{Secondary contributions}

  The contribution from the term
  \begin{equation}
   \frac{a^3\epsilon}{2c_\mathrm{s}^2}\frac{\mathrm{d}}{\mathrm{d}t}\Big(\frac{\eta}{c_\mathrm{s}^2}\Big)\zeta^2\dot\zeta \nonumber
  \end{equation}
  in equation\ \eqref{eq:S3int} is exactly zero since $\eta=0$ during the matter contraction.
  We can also neglect the contribution from the term
  \begin{equation}
   \frac{a\epsilon}{c_\mathrm{s}^2}(\epsilon-2s+1-c_\mathrm{s}^2)\zeta(\partial\zeta)^2 \nonumber
  \end{equation}
  since the leading order term of the resulting bispectrum is proportional to $c_\mathrm{s}^2k_i^2(\tau_B-\tilde{\tau}_B)^2$,
  which means that it is suppressed outside the sound Hubble radius.

\vspace*{0.5cm}

The above results differ from the ones of general single field inflation. As pointed out in \cite{Cai:2009fn},
two main reasons account for the different non-Gaussianities between matter bounce cosmology and inflation.
First, here the ``slow-roll'' parameter $\epsilon$ is of order one rather than being close to zero,
so the amplitudes are larger and the higher-order terms in $\epsilon$ are not suppressed.
Second, curvature perturbations grow on super-Hubble scales in a matter-dominated contracting universe, and this behaviour manifests itself
in the integral of equation\ \eqref{inin},
while for inflation, $\zeta$ usually remains constant after horizon-exit, so there is no such contribution.

In what follows, we summarize the above results and give a detailed analysis of the bispectrum.
In particular, the differences with the canonical single field matter bounce scenario are discussed.

\subsection{Summary of results}\label{sec:Shape}

One can gather all the contributions above and get the total shape function,
\begin{align}
\label{eq:Atot}
 \mathcal{A}_\mathrm{tot}=&~\left(-\frac{105}{32}+\frac{39}{16c_\mathrm{s}^2}+
 \frac{9c_\mathrm{s}^2}{128}\right)\sum_{i}k_i^3 +\frac{3}{256}(3c_\mathrm{s}^2+6)\sum_{i\neq j}k_i^2k_j+\frac{3}{256\prod_i k_i^2} \nonumber \\
 &\times\left[3c_\mathrm{s}^2\sum_i k_i^9
 +(10-9c_\mathrm{s}^2)\sum_{i\neq j}k_i^7k_j^2
 -(3c_\mathrm{s}^2+6)\sum_{i\neq j}k_i^6k_j^3+(9c_\mathrm{s}^2-4)\sum_{i\neq j}k_i^5k_j^4 \right]~,
\end{align}
where we have used $\epsilon=3/2$ and $\lambda/\Sigma=(1-c_\mathrm{s}^2)/{6c_\mathrm{s}^2}$ for the matter contraction stage.
Now the only free parameter in the total shape function is the sound speed $c_\mathrm{s}$.
In what follows, we shall discuss several interesting aspects of this result.

\subsubsection{Amplitude}

The size of non-Gaussianity is depicted by the dimensionless amplitude parameter
\begin{equation}
 f_{\rm NL}(\vec{k}_1,\vec{k}_2,\vec{k}_3)=\frac{10}{3}\frac{\mathcal{A}_\mathrm{tot}(\vec{k}_1,\vec{k}_2,\vec{k}_3)}{\sum_i k_i^3}~.
\end{equation}
As one can see in equation\ \eqref{eq:Atot}, for most values of $c_\mathrm{s}\in(0,1]$,
the first term dominates the total shape function, and roughly, $\fnl$ becomes
\be
\fnl\simeq -\frac{175}{16}+\frac{65}{8c_\mathrm{s}^2}+
 \frac{15c_\mathrm{s}^2}{64}~,
\ee
which yields $\fnl<0$ for $0.87\lesssim c_\mathrm{s}\leq1$ and $\fnl>0$ for $c_\mathrm{s}\lesssim 0.87$.
Thus, besides the negative amplitude in the canonical case \cite{Cai:2009fn}, a small sound speed in matter bounce cosmology
can produce a positive $\fnl$. In the next section, we shall further discuss its behaviour in different limits to confront observations.

\subsubsection{Shape}

\begin{figure}[tbhp]
\centering
\includegraphics[width=0.5\linewidth]{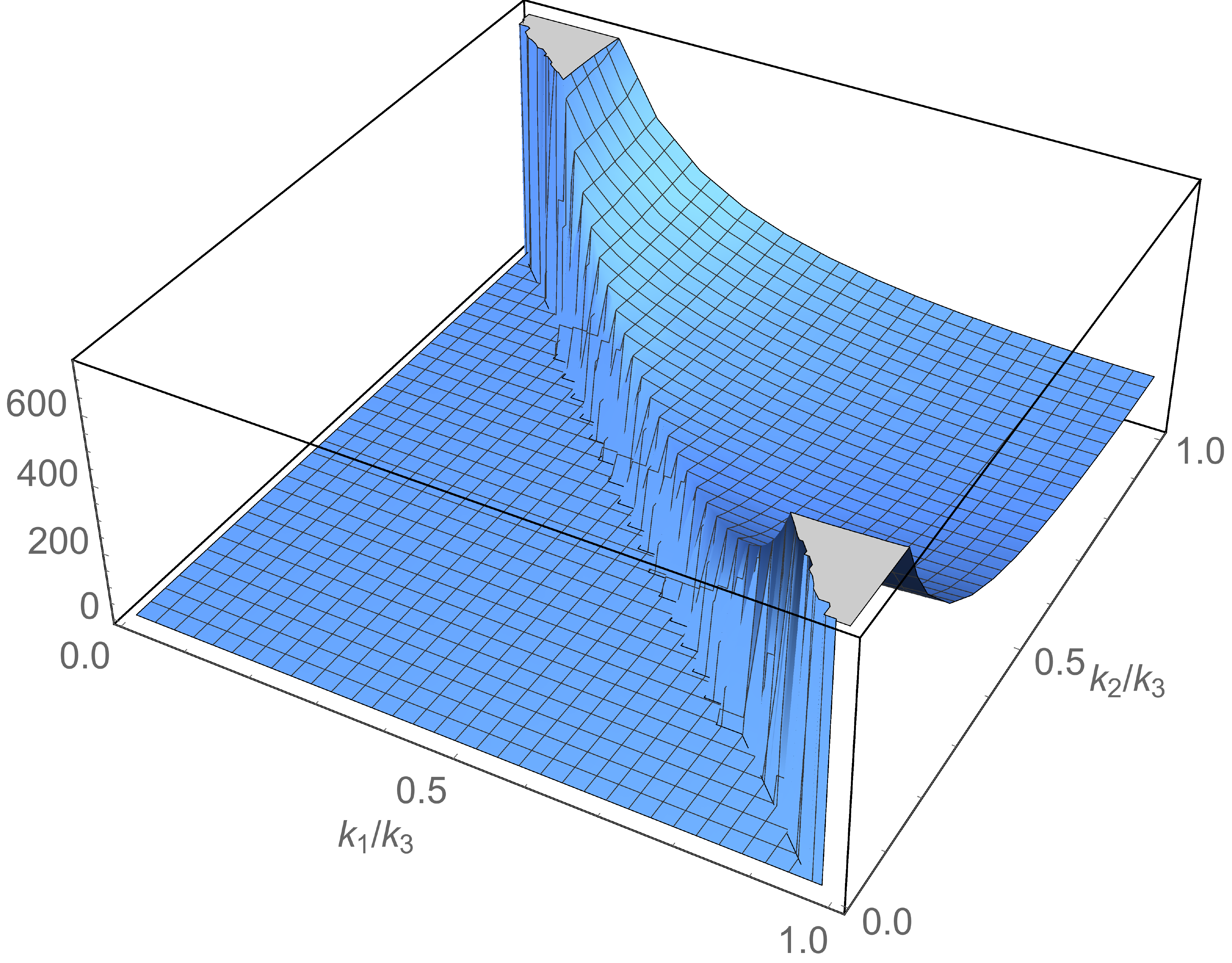}\includegraphics[width=0.5\linewidth]{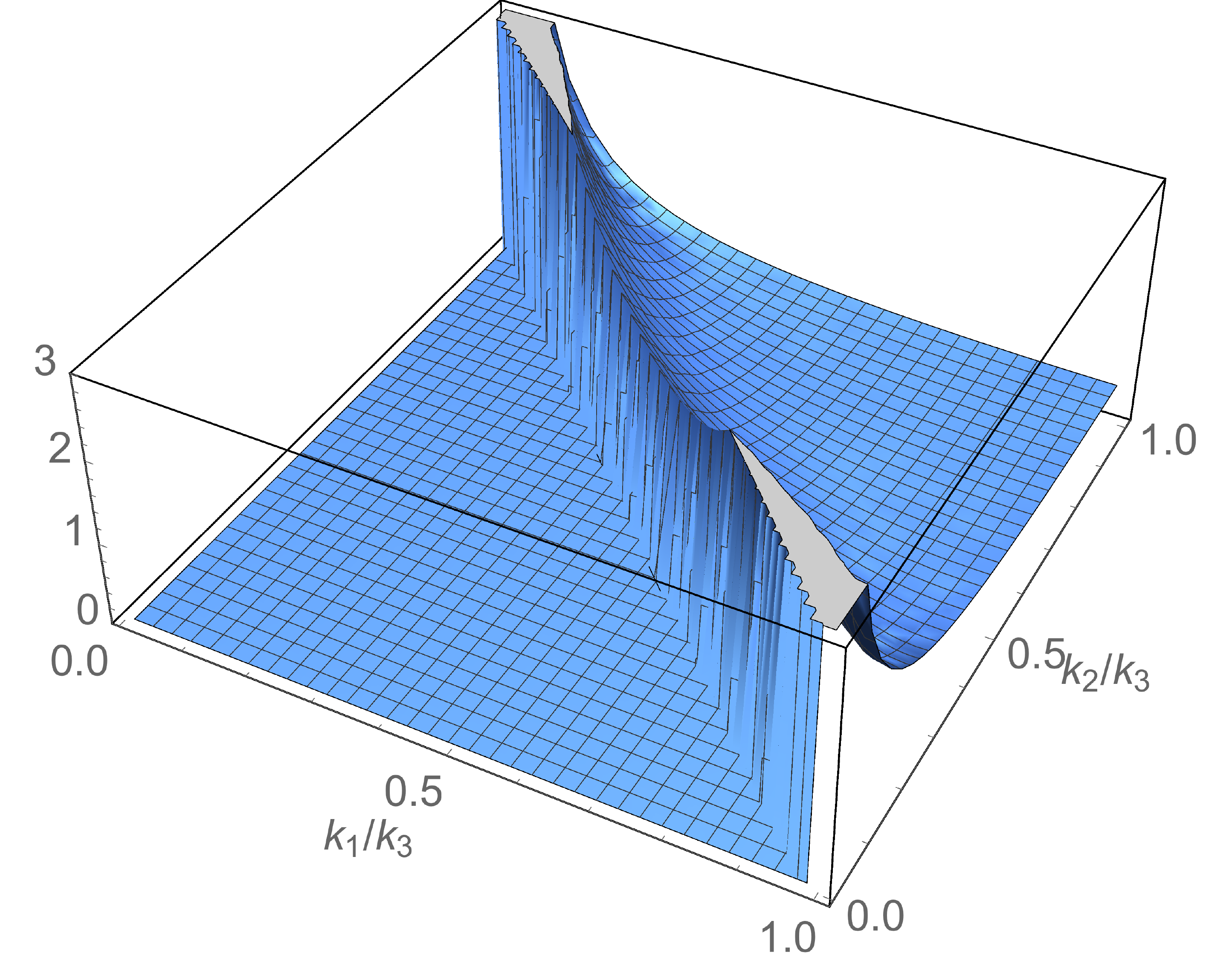}
\caption{The shape of $\mathcal{F}(k_1/k_3,k_2/k_3)$ for $c_\mathrm{s}=0.2$ (left panel) and $c_\mathrm{s}=0.87$ (right panel).}
\label{fig:shape}
\end{figure}

The shape of non-Gaussianity is described by the dimensionless shape function
\begin{equation}
\mathcal{F}(k_1/k_3,k_2/k_3)=\frac{\mathcal{A}_\mathrm{tot}}{k_1k_2k_3}~.
\end{equation}
Then, the first term in equation\ \eqref{eq:Atot} gives exactly the form of the local shape.
Thus, when the prefactor of the first term is nonvanishing ($c_\mathrm{s}\not\approx 0.87$),
the shape function is dominated by the local form, while the remaining terms just give some corrections.
The total shape of non-Gaussianity is shown in the left panel of Figure\ \ref{fig:shape},
which looks very similar to the plots in\ \cite{Cai:2009fn} for the canonical matter bounce
except that the amplitude is much larger here with $c_\mathrm{s}$ small.

At the same time, this result differs from the one of general single field inflation, where the equilateral form dominates the shape of
non-Gaussianity for $c_\mathrm{s}\ll 1$ \cite{Chen:2006nt}.
This is mainly caused by the different generation mechanisms of non-Gaussianity in these two scenarios.
For the matter bounce scenario, the growth of curvature perturbations after
Hubble radius exit makes a significant contribution to the final bispectrum.
Meanwhile, the local form is usually thought to be generated on super-Hubble scales since ``local''
means that the non-Gaussianity at one place is {\it disconnected} with the one at other places.
For general single field inflation, the dominant contribution is due to the enhanced interaction at horizon-crossing.
Thus, these two scenarios behave quite differently with a small sound speed.

It is also interesting to note that for $c_\mathrm{s}\approx 0.87$, the first term in equation\ \eqref{eq:Atot} vanishes,
so the shape function is dominated by the remaining terms. The shape of non-Gaussianity is plotted in the right panel of
Figure\ \ref{fig:shape} for this case, which is a new form different from the local one.
To the best of our knowledge, no other scenario can give rise to such a kind of shape, thus it can be seen as a distinguishable signature of
matter bounce cosmology for probes of non-Gaussianity.

\subsubsection{The squeezed limit}

Usually people are interested in the squeezed limit of the bispectrum ($k_1\ll k_2=k_3=k$), since its scaling behaviour is helpful for
clarifying the shapes of non-Gaussianity analytically.
Here in the squeezed limit ($k_1/k\rightarrow 0$), the dimensionless shape function can be expanded as
\be
\mathcal{F}(k_1/k_3,k_2/k_3)\simeq\frac{3}{8}\(-\frac{33}{2}+\frac{13}{c_\mathrm{s}^2}\)\frac{k}{k_1}
+\frac{3}{64}\(1+6c_\mathrm{s}^2\)\frac{k_1}{k}+\mathcal{O}\left(\left(\frac{k_1}{k}\right)^2\right)~.
\ee
The leading order term gives the scaling $ \mathcal{F}\sim k/k_1$ and
\be
\langle\zeta_{\vec{k}_1}\zeta_{\vec{k}_2}\zeta_{\vec{k}_3}\rangle_{\rm squeezed}\sim \frac{1}{k_1^3}~,
\ee
which is consistent with the dominant local form.
The only exception is when the coefficient of the first term vanishes ($c_\mathrm{s}=\sqrt{26/33}$)
and another scaling, $\mathcal{F}\sim k_1/k$, follows from the next-to-leading order term.

\section{Amplitude parameter of non-Gaussianities and implication for the no-go theorem}\label{sec:fNLandNoGo}

There are three forms of the amplitude parameter $\fnl$ that are of particular interest for cosmological observations.
They are called the ``local form'', the ``equilateral form'', and the ``folded form''.
The local form requires that one of the three momentum modes exits
the Hubble radius much earlier than the other two, e.g., $k_1\ll k_2=k_3$. In this limit, one can simplify the total shape function,
equation\ \eqref{eq:Atot}, to find
\begin{equation}
 f_\mathrm{NL}^\mathrm{local}\simeq-\frac{165}{16}+\frac{65}{8c_\mathrm{s}^2}~.
\end{equation}
The equilateral form requires that the three momenta form an equilateral triangle, i.e.~$k_1 = k_2 = k_3$. In this case, we obtain
\begin{equation}
 f_\mathrm{NL}^\mathrm{equil}\simeq-\frac{335}{32}+\frac{65}{8c_\mathrm{s}^2}+\frac{45c_\mathrm{s}^2}{128}~.
\end{equation}
The folded form has $k_1 = 2k_2 = 2k_3$, hence
\begin{equation}
 f_\mathrm{NL}^\mathrm{folded}\simeq-\frac{37}{4}+\frac{65}{8c_\mathrm{s}^2}~.
\end{equation}
As a result, in the limit where $c_\mathrm{s}^2\ll 1$, we find that
\begin{equation}
 f_\mathrm{NL}^\mathrm{local}\approx f_\mathrm{NL}^\mathrm{equil}\approx f_\mathrm{NL}^\mathrm{folded}\approx\frac{65}{8c_\mathrm{s}^2}\gg 1~.
\end{equation}
Let us recall from section\ \ref{sec:2ptfunction} that in order to satisfy the observational bound on the tensor-to-scalar
ratio, we must impose $c_\mathrm{s}\lesssim 0.0029$. This immediately implies
\begin{equation}
 f_\mathrm{NL}^\mathrm{local}\approx f_\mathrm{NL}^\mathrm{equil}\approx f_\mathrm{NL}^\mathrm{folded}\gtrsim 9.55\times 10^5\gg 1~.
\end{equation}
This amplitude of primordial non-Gaussianity is clearly ruled out according to the observations\ \cite{Ade:2015ava},
\begin{equation}
 f_\mathrm{NL}^\mathrm{local}=0.8\pm 5.0~,\ \ f_\mathrm{NL}^\mathrm{equil}=-4\pm 43~,\ \ f_\mathrm{NL}^\mathrm{ortho}=-26\pm 21~,
\end{equation}
thus ruling out the viability of the class of models studied here.

Alternatively, if one requires that, e.g., $-9.2\lesssim f_\mathrm{NL}^\mathrm{local}\lesssim 10.8$ (i.e., imposing $f_\mathrm{NL}^\mathrm{local}$
to be within the measured $2\sigma$ error bars), then one would need\footnote{Note that this constraint does not
exclude $c_\mathrm{s}\approx 0.87$, for which the new shape of non-Gaussianity in the right panel of Figure \ref{fig:shape} emerges.}
$c_\mathrm{s}\gtrsim 0.62$. However, this lower bound on the sound speed yields a tensor-to-scalar
ratio $r\gtrsim 14.88$, which is again clearly ruled out by observations\ \cite{Array:2015xqh}.

In summary, there is no region of parameter space where $c_\mathrm{s}$ can give a good, small tensor-to-scalar ratio (i.e., of order $0.1$ at most)
and good, small non-Gaussianities (i.e., of order 10 at most).
Therefore, independent of what happens during the bounce, we extend the no-go theorem conjectured in\ \cite{Quintin:2015rta}
to the following one:
\begin{theorem}
\textit{For quantum fluctuations generated during a matter-dominated
contracting phase, an upper bound on the tensor-to-scalar ratio ($r$)
is equivalent to a lower bound on the amount of primordial non-Gaussianities ($f_{\mathrm{NL}}$).
Furthermore, if
\begin{itemize}
 \item the matter contraction phase is due to a single (not necessarily canonical) scalar field,
 \item the same single scalar field allows for the violation of the NEC to produce a nonsingular bounce,
 \item and General Relativity holds at all energy scales,
\end{itemize}
then satisfying the current observational
upper bound on the tensor-to-scalar ratio cannot be done without contradicting the current observational upper bounds on
$f_{\mathrm{NL}}$ (and vice versa).}
\end{theorem}

\section{Conclusions and discussion}\label{sec:conclusion}

In this paper, we computed the two- and three-point correlation functions produced by
a generic $k$-essence scalar field in a matter-dominated contracting universe.
Comparing the power spectra of scalar and tensor modes, we found that the tensor-to-scalar ratio can
be appropriately suppressed if the sound speed associated with the $k$-essence
scalar field is sufficiently small.
In turn, we showed that the amplitude of the bispectrum is amplified by the smallness of the sound
speed\footnote{With a small sound speed, one may also reach the strong coupling regime
where the perturbative analysis breaks down.
This is known as the strong coupling problem\ \cite{Baumann:2011dt,Joyce:2011kh}, which affects many non-inflationary scenarios
(see in particular Appendix C of\ \cite{Baumann:2011dt}, which focuses on non-attractor models).
It represents a general independent theoretical constraint, but in the context of the matter bounce scenario, our no-go theorem
is more constraining due to current observational bounds.}.
As a result, it seems incompatible to suppress the tensor-to-scalar ratio below current observational bounds
without producing excessive non-Gaussianities.
This leads us to extend the conjecture of the no-go theorem, which effectively rules out a large class of nonsingular matter bounce models.

Although this seriously constrains nonsingular matter bounce cosmology as a viable alternative scenario to inflation,
there remain several classes of models that are not affected by this no-go theorem.
Indeed, one could still evade the no-go theorem assuming certain modified gravity models
as stated in\ \cite{Quintin:2015rta} (see references therein)
or with the introduction of one or several new fields.
For example, in the matter bounce curvaton scenario\ \cite{Cai:2011zx} (see also\ \cite{Qiu:2011cy,Alexander:2014uaa,Addazi:2016rnz}
for other nonsingular bouncing models using the curvaton mechanism)
and in the two-field matter bounce scenario\ \cite{Cai:2013kja},
entropy modes are generated by the presence of an additional scalar field, which are then converted to curvature perturbations.
In both models near the bounce, the kinetic term of the entropy field varies rapidly,
which acts as a tachyonic-like mass that amplifies (in a controlled way) the entropy fluctuations
while not affecting the tensor modes. As a result, the tensor-to-scalar ratio is suppressed (see\ \cite{Cai:2014xxa,Cai:2014bea}
for reviews of this process).
Furthermore, the production of non-Gaussianities in the matter bounce curvaton scenario has been estimated in\ \cite{Cai:2011zx},
and it indicated that sizable, negative non-Gaussianities appeared, yet still in agreement with current observations.
Accordingly, such a curvaton scenario does not appear to suffer from a no-go theorem.
However, there still remains to do an appropriate extensive analysis of the production of non-Gaussianities
when general multifields are included in the matter bounce scenario.

A similar curvaton mechanism is used in the new Ekpyrotic model\ \cite{Lehners:2007ac,Buchbinder:2007ad}
(extended in\ \cite{Qiu:2013eoa,Li:2013hga,Li:2014yla,Wilson-Ewing:2013bla}), which generates a nearly scale-invariant
power spectrum of curvature perturbations. In this case, however, the smallness of the observed tensor-to-scalar
ratio must be attributed to the fact that the tensor modes have a blue power spectrum when they exit the Hubble radius in a contracting phase
with $w\gg 1$. The new Ekpyrotic model originally predicted large
non-Gaussianities\ \cite{Buchbinder:2007tw,Buchbinder:2007at,Lehners:2007wc,Lehners:2008my,Lehners:2009ja}
(see also the reviews\ \cite{Lehners:2008vx,Lehners:2010fy}), but some more recent extensions can resolve
this issue\ \cite{Fertig:2013kwa,Ijjas:2014fja,Levy:2015awa,Fertig:2015ola,Fertig:2016czu}.
Thus, here as well, it appears that these types of models do not suffer from a similar no-go
theorem\footnote{Furthermore, Ekpyrotic models are robust against the growth of anisotropies in
a contracting universe. This is another challenge with the matter bounce scenario (see\ \cite{Cai:2013vm,Levy:2016xcl})
that will have to be overcome to have a viable theory.}.

We note that one might be able to prove the no-go conjecture of this paper borrowing similar techniques to
the effective field theory of inflation\ \cite{Cheung:2007st},
i.e.~by constructing an effective field theory of nonsingular bouncing cosmology (e.g., see the recent work of\ \cite{Cai:2016thi,Creminelli:2016zwa}).
In complete generality, this could allow us to find the exact and explicit
relation between the tensor-to-scalar ratio (which involves the power spectra of curvature and tensor modes)
and the bispectrum. In fact, the goal would be to find a consistency relation
for the three-point function in single field nonsingular bouncing cosmology
similar to what has been done in inflation\ \cite{Maldacena:2002vr,Creminelli:2004yq,Cheung:2007sv}.
This will be explored in a follow-up study.

Finally, we would like to emphasize that, for matter bounce cosmology, although the simplest $k$-essence model
is ruled out by the no-go theorem, the bispectrum with $c_\mathrm{s}\neq 1$ (as an independent result of this paper)
remains to be a probable target for future probes of non-Gaussianity.
This possibility relies on the aforementioned bouncing models that can evade the no-go theorem with other mechanisms.
In those cases, a nontrivial sound speed may still lead to the same behaviour of non-Gaussianities found in this paper, which potentially
can be detected by future observations.
Particularly, we predict a new shape with an amplitude still consistent with current observational limits, which can serve as
the distinctive signature of matter bounce cosmology and help us distinguish it from other very early universe theories.

\acknowledgments
We are grateful to Robert Brandenberger, Ziwei Wang, and Edward Wilson-Ewing for valuable comments and helpful discussions.
YFC, YBL, and DGW are supported in part 
by the Chinese National Youth Thousand Talents Program (No.~KJ2030220006),
by the USTC start-up funding (No.~KY2030000049),
by the National Natural Science Foundation of China (NSFC) (Nos.~11421303, 11653002),
and by the Fund for Fostering Talents in Basic Science of the NSFC (No.~J1310021).
JQ acknowledges financial support from the Walter C.~Sumner Memorial Fellowship and from the
Vanier Canada Graduate Scholarship administered by the Natural Sciences and Engineering Research Council of Canada (NSERC).
JQ also wishes to thank USTC for hospitality when this work was initiated.
DGW is also supported by a de Sitter Fellowship of the Netherlands Organization for Scientific Research (NWO).
Part of the numerical computations were done on the computer cluster LINDA in the particle cosmology group at USTC.

\appendix

\section{The ratio $\lambda/\Sigma$}\label{sec:ratiolambdasigma}

Let us recall the definition of $\Sigma$ and $\lambda$ in equations\ \eqref{eq:Sigmadef} and\ \eqref{eq:lambdadef}.
Their ratio is thus given by
\begin{equation}
 \frac{\lambda}{\Sigma}=\frac{1}{3}\left(X\frac{\Sigma_{,X}}{\Sigma}-1\right)~.
\end{equation}
Recalling the definition of $c_\mathrm{s}^2$ in equation\ \eqref{eq:defcs2s}, we note that
\begin{equation}
 \Sigma=X(P_{,X}+2XP_{,XX})=X\frac{P_{,X}}{c_\mathrm{s}^2}~.
\end{equation}
Also, recalling the expression for $\rho$ and $p$ in equation\ \eqref{eq:defrhop}, we find that
$2XP_{,X}=\rho+p$, and so, the above expression for $\Sigma$ becomes
\begin{equation}
 \Sigma=\frac{\rho+p}{2c_\mathrm{s}^2}~.
\end{equation}
Consequently,
\begin{equation}
\label{eq:XSigmaXoSigma1}
 X\frac{\Sigma_{,X}}{\Sigma}=X\frac{\rho_{,X}+p_{,X}}{\rho+p}-2X\frac{c_{\mathrm{s},X}}{c_\mathrm{s}}~.
\end{equation}

Working in the limit where $p=0$,
we note that $\rho=2XP_{,X}$, and so, $p_{,X}=P_{,X}=\rho/(2X)$,
which implies that $p_{,X}/\rho=1/(2X)$.
Also, $\rho_{,X}=p_{,X}/c_\mathrm{s}^2$ from the definition of the sound speed, and thus,
\begin{equation}
 \frac{\rho_{,X}}{\rho}=\frac{p_{,X}}{\rho c_\mathrm{s}^2}=\frac{1}{2c_\mathrm{s}^2X}~.
\end{equation}
Therefore, equation\ \eqref{eq:XSigmaXoSigma1} in the limit where $p=0$ becomes
\begin{equation}
\label{eq:ratio1}
 X\frac{\Sigma_{,X}}{\Sigma}=\frac{1}{2c_\mathrm{s}^2}+\frac{1}{2}-2X\frac{c_{\mathrm{s},X}}{c_\mathrm{s}}~.
\end{equation}

Alternatively, one can evaluate the ratio $\lambda/\Sigma$ as
\begin{equation}
\label{eq:lambdaoSigma}
 \frac{\lambda}{\Sigma}=\frac{1}{3}\left(\frac{\Sigma_{,X}}{\Sigma}X-1\right)=\frac{1}{3}\left(\frac{\dot\Sigma}{\Sigma}\frac{X}{\dot X}-1\right)~.
\end{equation}
Since we can write $\Sigma=H^2M_\mathrm{Pl}^2\epsilon/c_\mathrm{s}^2$
and recalling the definition of the slow-roll parameters in section\ \ref{sec:setup},
we get
\begin{equation}
\label{eq:SigmadotoHSigma}
 \frac{\dot\Sigma}{H\Sigma}=-2\epsilon+\eta-2s~.
\end{equation}
Now, we note that we can write
\begin{equation}
 \eta=\frac{\dot\epsilon}{H\epsilon}=\frac{\ddot H}{H\dot H}-2\frac{\dot H}{H^2}=\frac{\ddot H}{H\dot H}+2\epsilon~.
\end{equation}
Also, the Friedmann equation $M_\mathrm{Pl}^2\dot H=-XP_{,X}$ implies that
\begin{equation}
 \frac{\ddot H}{H\dot H}=\frac{1}{H}\left(\frac{\dot X}{X}+\frac{\dot P_{,X}}{P_{,X}}\right)~,
\end{equation}
and so,
\begin{equation}
\label{eq:dotXoHX}
 \frac{\dot X}{HX}=\eta-2\epsilon-\frac{\dot P_{,X}}{P_{,X}}~.
\end{equation}
Therefore, combining equation\ \eqref{eq:SigmadotoHSigma} and the above yields
\begin{equation}
 \frac{\dot\Sigma}{\Sigma}\frac{X}{\dot X}=\frac{-2\epsilon+\eta-2s}{-2\epsilon+\eta-\frac{\dot P_{,X}}{P_{,X}}}~.
\end{equation}
In the limit where $p=0$, we recall that $\epsilon=3/2$ and $\eta=0$, and as a result,
\begin{equation}
\label{eq:dotSigmaoSigmaXodotX}
 \frac{\dot\Sigma}{\Sigma}\frac{X}{\dot X}=\frac{3+2s}{3+\frac{\dot P_{,X}}{P_{,X}}}~.
\end{equation}
Comparing the above with equation\ \eqref{eq:ratio1}, since $(\dot\Sigma/\Sigma)(X/\dot X)=X\Sigma_{,X}/\Sigma$, we find
\begin{equation}
\label{eq:ratiotemp2}
 \frac{3+2s}{3+\frac{\dot P_{,X}}{P_{,X}}}=\frac{1}{2c_\mathrm{s}^2}+\frac{1}{2}-2X\frac{c_{\mathrm{s},X}}{c_\mathrm{s}}~,
\end{equation}
but
\begin{equation}
 -2X\frac{c_{\mathrm{s},X}}{c_\mathrm{s}}=-2\frac{X}{\dot X}\frac{\dot c_\mathrm{s}}{c_\mathrm{s}}=-2s\frac{HX}{\dot X}
 =\frac{-2s}{\eta-2\epsilon-\frac{\dot P_{,X}}{P_{,X}}}~,
\end{equation}
where the last equality follows from equation\ \eqref{eq:dotXoHX}.
Thus, equation\ \eqref{eq:ratiotemp2}, with $\epsilon=3/2$ and $\eta=0$, leaves us with
\begin{equation}
 \frac{3}{3+\frac{\dot P_{,X}}{P_{,X}}}=\frac{1}{2c_\mathrm{s}^2}+\frac{1}{2}~,
\end{equation}
and consequently,
\begin{equation}
 \frac{\dot P_{,X}}{P_{,X}}=-3\left(\frac{1-c_\mathrm{s}^2}{1+c_\mathrm{s}^2}\right)~.
\end{equation}
As a result, equation\ \eqref{eq:dotSigmaoSigmaXodotX} becomes
\begin{equation}
 \frac{\dot\Sigma}{\Sigma}\frac{X}{\dot X}=X\frac{\Sigma_{,X}}{\Sigma}=\frac{1}{2c_\mathrm{s}^2}\left(1+\frac{2}{3}s\right)(1+c_\mathrm{s}^2)~,
\end{equation}
and in the end,\ \eqref{eq:lambdaoSigma} is equivalent to
\begin{equation}
 \frac{\lambda}{\Sigma}=\frac{1}{3}\left[\frac{1}{2c_\mathrm{s}^2}\left(1+\frac{2}{3}s\right)(1+c_\mathrm{s}^2)-1\right]~.
\end{equation}
In the limit where $|s|\ll 1$, this reduces to
\begin{equation}
\label{eq:lambdaoSigmaresult}
 \frac{\lambda}{\Sigma}\simeq\frac{1}{3}\left[\frac{1+c_\mathrm{s}^2}{2c_\mathrm{s}^2}-1\right]=\frac{1-c_\mathrm{s}^2}{6c_\mathrm{s}^2}~.
\end{equation}
In comparison, DBI inflation has $\lambda/\Sigma=(1-c_\mathrm{s}^2)/(2c_\mathrm{s}^2)$ (see\ \cite{Chen:2006nt}).


\begin{thebibliography}{999}

%\cite{Brandenberger:2012zb}
\bibitem{Brandenberger:2012zb}
  R.~H.~Brandenberger,
  ``The Matter Bounce Alternative to Inflationary Cosmology,''
  arXiv:1206.4196 [astro-ph.CO].
  %%CITATION = ARXIV:1206.4196;%%

%\cite{Brandenberger:2011gk}
\bibitem{Brandenberger:2011gk}
  R.~H.~Brandenberger,
  ``Introduction to Early Universe Cosmology,''
  PoS ICFI {\bf 2010}, 001 (2010)
  [arXiv:1103.2271 [astro-ph.CO]].
  %%CITATION = ARXIV:1103.2271;%%

%\cite{Wands:1998yp}
\bibitem{Wands:1998yp}
  D.~Wands,
  ``Duality invariance of cosmological perturbation spectra,''
  Phys.\ Rev.\ D {\bf 60}, 023507 (1999)
  %doi:10.1103/PhysRevD.60.023507
  [gr-qc/9809062].
  %%CITATION = doi:10.1103/PhysRevD.60.023507;%%

%\cite{Finelli:2001sr}
\bibitem{Finelli:2001sr}
  F.~Finelli and R.~Brandenberger,
  ``On the generation of a scale invariant spectrum of adiabatic fluctuations in cosmological models with a contracting phase,''
  Phys.\ Rev.\ D {\bf 65}, 103522 (2002)
  %doi:10.1103/PhysRevD.65.103522
  [hep-th/0112249].
  %%CITATION = doi:10.1103/PhysRevD.65.103522;%%

%\cite{Martin:2000xs}
\bibitem{Martin:2000xs}
  J.~Martin and R.~H.~Brandenberger,
  ``The TransPlanckian problem of inflationary cosmology,''
  Phys.\ Rev.\ D {\bf 63}, 123501 (2001)
  %doi:10.1103/PhysRevD.63.123501
  [hep-th/0005209].
  %%CITATION = doi:10.1103/PhysRevD.63.123501;%%

%\cite{Wan:2015hya}
\bibitem{Wan:2015hya}
  Y.~Wan, T.~Qiu, F.~P.~Huang, Y.~F.~Cai, H.~Li and X.~Zhang,
  ``Bounce Inflation Cosmology with Standard Model Higgs Boson,''
  JCAP {\bf 1512}, no. 12, 019 (2015)
  %doi:10.1088/1475-7516/2015/12/019
  [arXiv:1509.08772 [gr-qc]].
  %%CITATION = doi:10.1088/1475-7516/2015/12/019;%%

%\cite{Borde:1993xh}
\bibitem{Borde:1993xh}
  A.~Borde and A.~Vilenkin,
  ``Eternal inflation and the initial singularity,''
  Phys.\ Rev.\ Lett.\  {\bf 72}, 3305 (1994)
  %doi:10.1103/PhysRevLett.72.3305
  [gr-qc/9312022].
  %%CITATION = doi:10.1103/PhysRevLett.72.3305;%%

%\cite{Borde:2001nh}
\bibitem{Borde:2001nh}
  A.~Borde, A.~H.~Guth and A.~Vilenkin,
  ``Inflationary space-times are incomplete in past directions,''
  Phys.\ Rev.\ Lett.\  {\bf 90}, 151301 (2003)
  %doi:10.1103/PhysRevLett.90.151301
  [gr-qc/0110012].
  %%CITATION = doi:10.1103/PhysRevLett.90.151301;%%

%\cite{Novello:2008ra}
\bibitem{Novello:2008ra}
  M.~Novello and S.~E.~P.~Bergliaffa,
  ``Bouncing Cosmologies,''
  Phys.\ Rept.\  {\bf 463}, 127 (2008)
  %doi:10.1016/j.physrep.2008.04.006
  [arXiv:0802.1634 [astro-ph]].
  %%CITATION = doi:10.1016/j.physrep.2008.04.006;%%

%\cite{Cai:2014bea}
\bibitem{Cai:2014bea}
  Y.~F.~Cai,
  ``Exploring Bouncing Cosmologies with Cosmological Surveys,''
  Sci.\ China Phys.\ Mech.\ Astron.\  {\bf 57}, 1414 (2014)
  %doi:10.1007/s11433-014-5512-3
  [arXiv:1405.1369 [hep-th]].
  %%CITATION = doi:10.1007/s11433-014-5512-3;%%

%\cite{Battefeld:2014uga}
\bibitem{Battefeld:2014uga}
  D.~Battefeld and P.~Peter,
  ``A Critical Review of Classical Bouncing Cosmologies,''
  Phys.\ Rept.\  {\bf 571}, 1 (2015)
  %doi:10.1016/j.physrep.2014.12.004
  [arXiv:1406.2790 [astro-ph.CO]].
  %%CITATION = doi:10.1016/j.physrep.2014.12.004;%%

%\cite{Brandenberger:2016vhg}
\bibitem{Brandenberger:2016vhg}
  R.~Brandenberger and P.~Peter,
  ``Bouncing Cosmologies: Progress and Problems,''
  Found.\ Phys.\ (2017)
  doi:10.1007/s10701-016-0057-0
  [arXiv:1603.05834 [hep-th]].
  %%CITATION = ARXIV:1603.05834;%%

%\cite{Nicolis:2008in}
\bibitem{Nicolis:2008in}
  A.~Nicolis, R.~Rattazzi and E.~Trincherini,
  ``The Galileon as a local modification of gravity,''
  Phys.\ Rev.\ D {\bf 79}, 064036 (2009)
  %doi:10.1103/PhysRevD.79.064036
  [arXiv:0811.2197 [hep-th]].
  %%CITATION = doi:10.1103/PhysRevD.79.064036;%%

%\cite{Horndeski:1974wa}
\bibitem{Horndeski:1974wa}
  G.~W.~Horndeski,
  ``Second-order scalar-tensor field equations in a four-dimensional space,''
  Int.\ J.\ Theor.\ Phys.\  {\bf 10}, 363 (1974).
  %doi:10.1007/BF01807638
  %%CITATION = doi:10.1007/BF01807638;%%

%\cite{Qiu:2011cy}
\bibitem{Qiu:2011cy}
  T.~Qiu, J.~Evslin, Y.~F.~Cai, M.~Li and X.~Zhang,
  ``Bouncing Galileon Cosmologies,''
  JCAP {\bf 1110}, 036 (2011)
  %doi:10.1088/1475-7516/2011/10/036
  [arXiv:1108.0593 [hep-th]].
  %%CITATION = doi:10.1088/1475-7516/2011/10/036;%%

%\cite{Easson:2011zy}
\bibitem{Easson:2011zy}
  D.~A.~Easson, I.~Sawicki and A.~Vikman,
  ``G-Bounce,''
  JCAP {\bf 1111}, 021 (2011)
  %doi:10.1088/1475-7516/2011/11/021
  [arXiv:1109.1047 [hep-th]].
  %%CITATION = doi:10.1088/1475-7516/2011/11/021;%%

%\cite{Cai:2012va}
\bibitem{Cai:2012va}
  Y.~F.~Cai, D.~A.~Easson and R.~Brandenberger,
  ``Towards a Nonsingular Bouncing Cosmology,''
  JCAP {\bf 1208}, 020 (2012)
  %doi:10.1088/1475-7516/2012/08/020
  [arXiv:1206.2382 [hep-th]].
  %%CITATION = doi:10.1088/1475-7516/2012/08/020;%%

%\cite{Cai:2013vm}
\bibitem{Cai:2013vm}
  Y.~F.~Cai, R.~Brandenberger and P.~Peter,
  ``Anisotropy in a Nonsingular Bounce,''
  Class.\ Quant.\ Grav.\  {\bf 30}, 075019 (2013)
  %doi:10.1088/0264-9381/30/7/075019
  [arXiv:1301.4703 [gr-qc]].
  %%CITATION = doi:10.1088/0264-9381/30/7/075019;%%

%\cite{Osipov:2013ssa}
\bibitem{Osipov:2013ssa}
  M.~Osipov and V.~Rubakov,
  ``Galileon bounce after ekpyrotic contraction,''
  JCAP {\bf 1311}, 031 (2013)
  %doi:10.1088/1475-7516/2013/11/031
  [arXiv:1303.1221 [hep-th]].
  %%CITATION = doi:10.1088/1475-7516/2013/11/031;%%

%\cite{Battarra:2014tga}
\bibitem{Battarra:2014tga}
  L.~Battarra, M.~Koehn, J.~L.~Lehners and B.~A.~Ovrut,
  ``Cosmological Perturbations Through a Non-Singular Ghost-Condensate/Galileon Bounce,''
  JCAP {\bf 1407}, 007 (2014)
  %doi:10.1088/1475-7516/2014/07/007
  [arXiv:1404.5067 [hep-th]].
  %%CITATION = doi:10.1088/1475-7516/2014/07/007;%%

%\cite{Ijjas:2016tpn}
\bibitem{Ijjas:2016tpn}
  A.~Ijjas and P.~J.~Steinhardt,
  ``Classically stable non-singular cosmological bounces,''
  Phys.\ Rev.\ Lett.\  {\bf 117}, no. 12, 121304 (2016)
  %doi:10.1103/PhysRevLett.117.121304
  [arXiv:1606.08880 [gr-qc]].
  %%CITATION = doi:10.1103/PhysRevLett.117.121304;%%

%\cite{Ijjas:2016vtq}
\bibitem{Ijjas:2016vtq}
  A.~Ijjas and P.~J.~Steinhardt,
  ``Fully stable cosmological solutions with a non-singular classical bounce,''
  Phys.\ Lett.\ B {\bf 764}, 289 (2017)
  %doi:10.1016/j.physletb.2016.11.047
  [arXiv:1609.01253 [gr-qc]].
  %%CITATION = doi:10.1016/j.physletb.2016.11.047;%%

%\cite{Libanov:2016kfc}
\bibitem{Libanov:2016kfc}
  M.~Libanov, S.~Mironov and V.~Rubakov,
  ``Generalized Galileons: instabilities of bouncing and Genesis cosmologies and modified Genesis,''
  JCAP {\bf 1608}, no. 08, 037 (2016)
  %doi:10.1088/1475-7516/2016/08/037
  [arXiv:1605.05992 [hep-th]].
  %%CITATION = doi:10.1088/1475-7516/2016/08/037;%%

%\cite{Kobayashi:2016xpl}
\bibitem{Kobayashi:2016xpl}
  T.~Kobayashi,
  ``Generic instabilities of nonsingular cosmologies in Horndeski theory: A no-go theorem,''
  Phys.\ Rev.\ D {\bf 94}, no. 4, 043511 (2016)
  %doi:10.1103/PhysRevD.94.043511
  [arXiv:1606.05831 [hep-th]].
  %%CITATION = doi:10.1103/PhysRevD.94.043511;%%

%\cite{Cai:2016thi}
\bibitem{Cai:2016thi}
  Y.~Cai, Y.~Wan, H.~G.~Li, T.~Qiu and Y.~S.~Piao,
  ``The Effective Field Theory of nonsingular cosmology,''
  JHEP {\bf 1701}, 090 (2017)
  %doi:10.1007/JHEP01(2017)090
  [arXiv:1610.03400 [gr-qc]].
  %%CITATION = doi:10.1007/JHEP01(2017)090;%%

%\cite{Creminelli:2016zwa}
\bibitem{Creminelli:2016zwa}
  P.~Creminelli, D.~Pirtskhalava, L.~Santoni and E.~Trincherini,
  ``Stability of Geodesically Complete Cosmologies,''
  JCAP {\bf 1611}, no. 11, 047 (2016)
  %doi:10.1088/1475-7516/2016/11/047
  [arXiv:1610.04207 [hep-th]].
  %%CITATION = doi:10.1088/1475-7516/2016/11/047;%%

%\cite{Lehners:2015mra}
\bibitem{Lehners:2015mra}
  J.~L.~Lehners and E.~Wilson-Ewing,
  ``Running of the scalar spectral index in bouncing cosmologies,''
  JCAP {\bf 1510}, no. 10, 038 (2015)
  %doi:10.1088/1475-7516/2015/10/038
  [arXiv:1507.08112 [astro-ph.CO]].
  %%CITATION = doi:10.1088/1475-7516/2015/10/038;%%

%\cite{Cai:2016hea}
\bibitem{Cai:2016hea}
  Y.~F.~Cai, A.~Marciano, D.~G.~Wang and E.~Wilson-Ewing,
  ``Bouncing cosmologies with dark matter and dark energy,''
  Universe {\bf 3}, no. 1, 1 (2016)
  %doi:10.3390/universe3010001
  [arXiv:1610.00938 [astro-ph.CO]].
  %%CITATION = doi:10.3390/universe3010001;%%

%\cite{Maldacena:2002vr}
\bibitem{Maldacena:2002vr}
  J.~M.~Maldacena,
  ``Non-Gaussian features of primordial fluctuations in single field inflationary models,''
  JHEP {\bf 0305}, 013 (2003)
  %doi:10.1088/1126-6708/2003/05/013
  [astro-ph/0210603].
  %%CITATION = doi:10.1088/1126-6708/2003/05/013;%%

%\cite{Chen:2010xka}
\bibitem{Chen:2010xka}
  X.~Chen,
  ``Primordial Non-Gaussianities from Inflation Models,''
  Adv.\ Astron.\  {\bf 2010}, 638979 (2010)
  %doi:10.1155/2010/638979
  [arXiv:1002.1416 [astro-ph.CO]].
  %%CITATION = doi:10.1155/2010/638979;%%

%\cite{Wang:2013eqj}
\bibitem{Wang:2013eqj}
  Y.~Wang,
  ``Inflation, Cosmic Perturbations and Non-Gaussianities,''
  Commun.\ Theor.\ Phys.\  {\bf 62}, 109 (2014)
  %doi:10.1088/0253-6102/62/1/19
  [arXiv:1303.1523 [hep-th]].
  %%CITATION = doi:10.1088/0253-6102/62/1/19;%%

%\cite{ArmendarizPicon:2000dh}
\bibitem{ArmendarizPicon:2000dh}
  C.~Armendariz-Picon, V.~F.~Mukhanov and P.~J.~Steinhardt,
  ``A Dynamical solution to the problem of a small cosmological constant and late time cosmic acceleration,''
  Phys.\ Rev.\ Lett.\  {\bf 85}, 4438 (2000)
  %doi:10.1103/PhysRevLett.85.4438
  [astro-ph/0004134].
  %%CITATION = doi:10.1103/PhysRevLett.85.4438;%%

%\cite{ArmendarizPicon:2000ah}
\bibitem{ArmendarizPicon:2000ah}
  C.~Armendariz-Picon, V.~F.~Mukhanov and P.~J.~Steinhardt,
  ``Essentials of $k$-essence,''
  Phys.\ Rev.\ D {\bf 63}, 103510 (2001)
  %doi:10.1103/PhysRevD.63.103510
  [astro-ph/0006373].
  %%CITATION = doi:10.1103/PhysRevD.63.103510;%%

%\cite{ArmendarizPicon:1999rj}
\bibitem{ArmendarizPicon:1999rj}
  C.~Armendariz-Picon, T.~Damour and V.~F.~Mukhanov,
  ``$k$-Inflation,''
  Phys.\ Lett.\ B {\bf 458}, 209 (1999)
  %doi:10.1016/S0370-2693(99)00603-6
  [hep-th/9904075].
  %%CITATION = doi:10.1016/S0370-2693(99)00603-6;%%

%\cite{Garriga:1999vw}
\bibitem{Garriga:1999vw}
  J.~Garriga and V.~F.~Mukhanov,
  ``Perturbations in $k$-inflation,''
  Phys.\ Lett.\ B {\bf 458}, 219 (1999)
  %doi:10.1016/S0370-2693(99)00602-4
  [hep-th/9904176].
  %%CITATION = doi:10.1016/S0370-2693(99)00602-4;%%

%\cite{Silverstein:2003hf}
\bibitem{Silverstein:2003hf}
  E.~Silverstein and D.~Tong,
  ``Scalar speed limits and cosmology: Acceleration from D-cceleration,''
  Phys.\ Rev.\ D {\bf 70}, 103505 (2004)
  %doi:10.1103/PhysRevD.70.103505
  [hep-th/0310221].
  %%CITATION = doi:10.1103/PhysRevD.70.103505;%%

%\cite{Alishahiha:2004eh}
\bibitem{Alishahiha:2004eh}
  M.~Alishahiha, E.~Silverstein and D.~Tong,
  ``DBI in the sky,''
  Phys.\ Rev.\ D {\bf 70}, 123505 (2004)
  %doi:10.1103/PhysRevD.70.123505
  [hep-th/0404084].
  %%CITATION = doi:10.1103/PhysRevD.70.123505;%%

%\cite{Chen:2006nt}
\bibitem{Chen:2006nt}
  X.~Chen, M.~x.~Huang, S.~Kachru and G.~Shiu,
  ``Observational signatures and non-Gaussianities of general single field inflation,''
  JCAP {\bf 0701}, 002 (2007)
  %doi:10.1088/1475-7516/2007/01/002
  [hep-th/0605045].
  %%CITATION = doi:10.1088/1475-7516/2007/01/002;%%

%\cite{Seery:2005wm}
\bibitem{Seery:2005wm}
  D.~Seery and J.~E.~Lidsey,
  ``Primordial non-Gaussianities in single field inflation,''
  JCAP {\bf 0506}, 003 (2005)
  %doi:10.1088/1475-7516/2005/06/003
  [astro-ph/0503692].
  %%CITATION = doi:10.1088/1475-7516/2005/06/003;%%

%\cite{Cheung:2007st}
\bibitem{Cheung:2007st}
  C.~Cheung, P.~Creminelli, A.~L.~Fitzpatrick, J.~Kaplan and L.~Senatore,
  ``The Effective Field Theory of Inflation,''
  JHEP {\bf 0803}, 014 (2008)
  %doi:10.1088/1126-6708/2008/03/014
  [arXiv:0709.0293 [hep-th]].
  %%CITATION = doi:10.1088/1126-6708/2008/03/014;%%

%\cite{Noller:2011hd}
\bibitem{Noller:2011hd}
  J.~Noller and J.~Magueijo,
  ``Non-Gaussianity in single field models without slow-roll,''
  Phys.\ Rev.\ D {\bf 83}, 103511 (2011)
  %doi:10.1103/PhysRevD.83.103511
  [arXiv:1102.0275 [astro-ph.CO]].
  %%CITATION = doi:10.1103/PhysRevD.83.103511;%%

%\cite{Cai:2009fn}
\bibitem{Cai:2009fn}
  Y.~F.~Cai, W.~Xue, R.~Brandenberger and X.~Zhang,
  ``Non-Gaussianity in a Matter Bounce,''
  JCAP {\bf 0905}, 011 (2009)
  %doi:10.1088/1475-7516/2009/05/011
  [arXiv:0903.0631 [astro-ph.CO]].
  %%CITATION = doi:10.1088/1475-7516/2009/05/011;%%

%\cite{Deffayet:2011gz}
\bibitem{Deffayet:2011gz}
  C.~Deffayet, X.~Gao, D.~A.~Steer and G.~Zahariade,
  ``From k-essence to generalised Galileons,''
  Phys.\ Rev.\ D {\bf 84}, 064039 (2011)
  %doi:10.1103/PhysRevD.84.064039
  [arXiv:1103.3260 [hep-th]].
  %%CITATION = doi:10.1103/PhysRevD.84.064039;%%

%\cite{Kobayashi:2010cm}
\bibitem{Kobayashi:2010cm}
  T.~Kobayashi, M.~Yamaguchi and J.~Yokoyama,
  ``G-inflation: Inflation driven by the Galileon field,''
  Phys.\ Rev.\ Lett.\  {\bf 105}, 231302 (2010)
  %doi:10.1103/PhysRevLett.105.231302
  [arXiv:1008.0603 [hep-th]].
  %%CITATION = doi:10.1103/PhysRevLett.105.231302;%%

%\cite{Burrage:2010cu}
\bibitem{Burrage:2010cu}
  C.~Burrage, C.~de Rham, D.~Seery and A.~J.~Tolley,
  ``Galileon inflation,''
  JCAP {\bf 1101}, 014 (2011)
  %doi:10.1088/1475-7516/2011/01/014
  [arXiv:1009.2497 [hep-th]].
  %%CITATION = doi:10.1088/1475-7516/2011/01/014;%%

%\cite{Creminelli:2010qf}
\bibitem{Creminelli:2010qf}
  P.~Creminelli, G.~D'Amico, M.~Musso, J.~Norena and E.~Trincherini,
  ``Galilean symmetry in the effective theory of inflation: new shapes of non-Gaussianity,''
  JCAP {\bf 1102}, 006 (2011)
  %doi:10.1088/1475-7516/2011/02/006
  [arXiv:1011.3004 [hep-th]].
  %%CITATION = doi:10.1088/1475-7516/2011/02/006;%%

%\cite{Kobayashi:2011nu}
\bibitem{Kobayashi:2011nu}
  T.~Kobayashi, M.~Yamaguchi and J.~Yokoyama,
  ``Generalized G-inflation: Inflation with the most general second-order field equations,''
  Prog.\ Theor.\ Phys.\  {\bf 126}, 511 (2011)
  %doi:10.1143/PTP.126.511
  [arXiv:1105.5723 [hep-th]].
  %%CITATION = doi:10.1143/PTP.126.511;%%

%\cite{Gao:2011qe}
\bibitem{Gao:2011qe}
  X.~Gao and D.~A.~Steer,
  ``Inflation and primordial non-Gaussianities of `generalized Galileons',''
  JCAP {\bf 1112}, 019 (2011)
  %doi:10.1088/1475-7516/2011/12/019
  [arXiv:1107.2642 [astro-ph.CO]].
  %%CITATION = doi:10.1088/1475-7516/2011/12/019;%%

%\cite{Cai:2008qw}
\bibitem{Cai:2008qw}
  Y.~F.~Cai, T.~t.~Qiu, R.~Brandenberger and X.~m.~Zhang,
  ``A Nonsingular Cosmology with a Scale-Invariant Spectrum of Cosmological Perturbations from Lee-Wick Theory,''
  Phys.\ Rev.\ D {\bf 80}, 023511 (2009)
  %doi:10.1103/PhysRevD.80.023511
  [arXiv:0810.4677 [hep-th]].
  %%CITATION = doi:10.1103/PhysRevD.80.023511;%%

%\cite{Cai:2014xxa}
\bibitem{Cai:2014xxa}
  Y.~F.~Cai, J.~Quintin, E.~N.~Saridakis and E.~Wilson-Ewing,
  ``Nonsingular bouncing cosmologies in light of BICEP2,''
  JCAP {\bf 1407}, 033 (2014)
  %doi:10.1088/1475-7516/2014/07/033
  [arXiv:1404.4364 [astro-ph.CO]].
  %%CITATION = doi:10.1088/1475-7516/2014/07/033;%%

%\cite{Quintin:2015rta}
\bibitem{Quintin:2015rta}
  J.~Quintin, Z.~Sherkatghanad, Y.~F.~Cai and R.~H.~Brandenberger,
  ``Evolution of cosmological perturbations and the production of non-Gaussianities through a nonsingular bounce: Indications for a no-go theorem in single field matter bounce cosmologies,''
  Phys.\ Rev.\ D {\bf 92}, no. 6, 063532 (2015)
  %doi:10.1103/PhysRevD.92.063532
  [arXiv:1508.04141 [hep-th]].
  %%CITATION = doi:10.1103/PhysRevD.92.063532;%%

%\cite{Array:2015xqh}
\bibitem{Array:2015xqh}
  P.~A.~R.~Ade {\it et al.} [BICEP2 and Keck Array Collaborations],
  ``Improved Constraints on Cosmology and Foregrounds from BICEP2 and Keck Array Cosmic Microwave Background Data with Inclusion of 95 GHz Band,''
  Phys.\ Rev.\ Lett.\  {\bf 116}, 031302 (2016)
  %doi:10.1103/PhysRevLett.116.031302
  [arXiv:1510.09217 [astro-ph.CO]].
  %%CITATION = doi:10.1103/PhysRevLett.116.031302;%%

%\cite{Xue:2013bva}
\bibitem{Xue:2013bva}
  B.~Xue, D.~Garfinkle, F.~Pretorius and P.~J.~Steinhardt,
  ``Nonperturbative analysis of the evolution of cosmological perturbations through a nonsingular bounce,''
  Phys.\ Rev.\ D {\bf 88}, 083509 (2013)
  %doi:10.1103/PhysRevD.88.083509
  [arXiv:1308.3044 [gr-qc]].
  %%CITATION = doi:10.1103/PhysRevD.88.083509;%%

%\cite{Cai:2014jla}
\bibitem{Cai:2014jla}
  Y.~F.~Cai and E.~Wilson-Ewing,
  ``A $\Lambda$CDM bounce scenario,''
  JCAP {\bf 1503}, no. 03, 006 (2015)
  %doi:10.1088/1475-7516/2015/03/006
  [arXiv:1412.2914 [gr-qc]].
  %%CITATION = doi:10.1088/1475-7516/2015/03/006;%%

%\cite{Wilson-Ewing:2015sfx}
\bibitem{Wilson-Ewing:2015sfx}
  E.~Wilson-Ewing,
  ``Separate universes in loop quantum cosmology: framework and applications,''
  Int.\ J.\ Mod.\ Phys.\ D {\bf 25}, no. 08, 1642002 (2016)
  %doi:10.1142/S0218271816420025
  [arXiv:1512.05743 [gr-qc]].
  %%CITATION = doi:10.1142/S0218271816420025;%%

%\cite{Gao:2014hea}
\bibitem{Gao:2014hea}
  X.~Gao, M.~Lilley and P.~Peter,
  ``Production of non-gaussianities through a positive spatial curvature bouncing phase,''
  JCAP {\bf 1407}, 010 (2014)
  %doi:10.1088/1475-7516/2014/07/010
  [arXiv:1403.7958 [gr-qc]].
  %%CITATION = doi:10.1088/1475-7516/2014/07/010;%%

%\cite{Gao:2014eaa}
\bibitem{Gao:2014eaa}
  X.~Gao, M.~Lilley and P.~Peter,
  ``Non-Gaussianity excess problem in classical bouncing cosmologies,''
  Phys.\ Rev.\ D {\bf 91}, no. 2, 023516 (2015)
  %doi:10.1103/PhysRevD.91.023516
  [arXiv:1406.4119 [gr-qc]].
  %%CITATION = doi:10.1103/PhysRevD.91.023516;%%

%\cite{Ade:2015ava}
\bibitem{Ade:2015ava}
  P.~A.~R.~Ade {\it et al.} [Planck Collaboration],
  ``Planck 2015 results. XVII. Constraints on primordial non-Gaussianity,''
  Astron.\ Astrophys.\  {\bf 594}, A17 (2016)
  %doi:10.1051/0004-6361/201525836
  [arXiv:1502.01592 [astro-ph.CO]].
  %%CITATION = doi:10.1051/0004-6361/201525836;%%

%\cite{Cai:2015vzv}
\bibitem{Cai:2015vzv}
  Y.~F.~Cai, F.~Duplessis, D.~A.~Easson and D.~G.~Wang,
  ``Searching for a matter bounce cosmology with low redshift observations,''
  Phys.\ Rev.\ D {\bf 93}, no. 4, 043546 (2016)
  %doi:10.1103/PhysRevD.93.043546
  [arXiv:1512.08979 [astro-ph.CO]].
  %%CITATION = doi:10.1103/PhysRevD.93.043546;%%

%\cite{Lin:2010pf}
\bibitem{Lin:2010pf}
  C.~Lin, R.~H.~Brandenberger and L.~Perreault Levasseur,
  ``A Matter Bounce By Means of Ghost Condensation,''
  JCAP {\bf 1104}, 019 (2011)
  %doi:10.1088/1475-7516/2011/04/019
  [arXiv:1007.2654 [hep-th]].
  %%CITATION = doi:10.1088/1475-7516/2011/04/019;%%

%\cite{Quintin:2016qro}
\bibitem{Quintin:2016qro}
  J.~Quintin and R.~H.~Brandenberger,
  ``Black hole formation in a contracting universe,''
  JCAP {\bf 1611}, no. 11, 029 (2016)
  %doi:10.1088/1475-7516/2016/11/029
  [arXiv:1609.02556 [astro-ph.CO]].
  %%CITATION = doi:10.1088/1475-7516/2016/11/029;%%

%\cite{Baumann:2011dt}
\bibitem{Baumann:2011dt}
  D.~Baumann, L.~Senatore and M.~Zaldarriaga,
  ``Scale-Invariance and the Strong Coupling Problem,''
  JCAP {\bf 1105}, 004 (2011)
  %doi:10.1088/1475-7516/2011/05/004
  [arXiv:1101.3320 [hep-th]].
  %%CITATION = doi:10.1088/1475-7516/2011/05/004;%%

%\cite{Joyce:2011kh}
\bibitem{Joyce:2011kh}
  A.~Joyce and J.~Khoury,
  ``Strong Coupling Problem with Time-Varying Sound Speed,''
  Phys.\ Rev.\ D {\bf 84}, 083514 (2011)
  %doi:10.1103/PhysRevD.84.083514
  [arXiv:1107.3550 [hep-th]].
  %%CITATION = doi:10.1103/PhysRevD.84.083514;%%

%\cite{Cai:2011zx}
\bibitem{Cai:2011zx}
  Y.~F.~Cai, R.~Brandenberger and X.~Zhang,
  ``The Matter Bounce Curvaton Scenario,''
  JCAP {\bf 1103}, 003 (2011)
  %doi:10.1088/1475-7516/2011/03/003
  [arXiv:1101.0822 [hep-th]].
  %%CITATION = doi:10.1088/1475-7516/2011/03/003;%%

%\cite{Alexander:2014uaa}
\bibitem{Alexander:2014uaa}
  S.~Alexander, Y.~F.~Cai and A.~Marciano,
  ``Fermi-bounce cosmology and the fermion curvaton mechanism,''
  Phys.\ Lett.\ B {\bf 745}, 97 (2015)
  %doi:10.1016/j.physletb.2015.04.026
  [arXiv:1406.1456 [gr-qc]].
  %%CITATION = doi:10.1016/j.physletb.2015.04.026;%%

%\cite{Addazi:2016rnz}
\bibitem{Addazi:2016rnz}
  A.~Addazi, S.~Alexander, Y.~F.~Cai and A.~Marciano,
  ``Dark matter and baryogenesis in the Fermi-bounce curvaton mechanism,''
  arXiv:1612.00632 [gr-qc].

%\cite{Cai:2013kja}
\bibitem{Cai:2013kja}
  Y.~F.~Cai, E.~McDonough, F.~Duplessis and R.~H.~Brandenberger,
  ``Two Field Matter Bounce Cosmology,''
  JCAP {\bf 1310}, 024 (2013)
  %doi:10.1088/1475-7516/2013/10/024
  [arXiv:1305.5259 [hep-th]].
  %%CITATION = doi:10.1088/1475-7516/2013/10/024;%%

%\cite{Lehners:2007ac}
\bibitem{Lehners:2007ac}
  J.~L.~Lehners, P.~McFadden, N.~Turok and P.~J.~Steinhardt,
  ``Generating ekpyrotic curvature perturbations before the big bang,''
  Phys.\ Rev.\ D {\bf 76}, 103501 (2007)
  %doi:10.1103/PhysRevD.76.103501
  [hep-th/0702153 [HEP-TH]].
  %%CITATION = doi:10.1103/PhysRevD.76.103501;%%

%\cite{Buchbinder:2007ad}
\bibitem{Buchbinder:2007ad}
  E.~I.~Buchbinder, J.~Khoury and B.~A.~Ovrut,
  ``New Ekpyrotic cosmology,''
  Phys.\ Rev.\ D {\bf 76}, 123503 (2007)
  %doi:10.1103/PhysRevD.76.123503
  [hep-th/0702154].
  %%CITATION = doi:10.1103/PhysRevD.76.123503;%%

%\cite{Qiu:2013eoa}
\bibitem{Qiu:2013eoa}
  T.~Qiu, X.~Gao and E.~N.~Saridakis,
  ``Towards anisotropy-free and nonsingular bounce cosmology with scale-invariant perturbations,''
  Phys.\ Rev.\ D {\bf 88} (2013) no.4,  043525
  %doi:10.1103/PhysRevD.88.043525
  [arXiv:1303.2372 [astro-ph.CO]].
  %%CITATION = doi:10.1103/PhysRevD.88.043525;%%

%\cite{Li:2013hga}
\bibitem{Li:2013hga}
  M.~Li,
  ``Note on the production of scale-invariant entropy perturbation in the Ekpyrotic universe,''
  Phys.\ Lett.\ B {\bf 724}, 192 (2013)
  %doi:10.1016/j.physletb.2013.06.035
  [arXiv:1306.0191 [hep-th]].
  %%CITATION = doi:10.1016/j.physletb.2013.06.035;%%

%\cite{Li:2014yla}
\bibitem{Li:2014yla}
  M.~Li,
  ``Entropic mechanisms with generalized scalar fields in the Ekpyrotic universe,''
  Phys.\ Lett.\ B {\bf 741}, 320 (2015)
  %doi:10.1016/j.physletb.2015.01.009
  [arXiv:1411.7626 [hep-th]].
  %%CITATION = doi:10.1016/j.physletb.2015.01.009;%%

%\cite{Wilson-Ewing:2013bla}
\bibitem{Wilson-Ewing:2013bla}
  E.~Wilson-Ewing,
  ``Ekpyrotic loop quantum cosmology,''
  JCAP {\bf 1308}, 015 (2013)
  %doi:10.1088/1475-7516/2013/08/015
  [arXiv:1306.6582 [gr-qc]].
  %%CITATION = doi:10.1088/1475-7516/2013/08/015;%%

%\cite{Buchbinder:2007tw}
\bibitem{Buchbinder:2007tw}
  E.~I.~Buchbinder, J.~Khoury and B.~A.~Ovrut,
  ``On the initial conditions in new ekpyrotic cosmology,''
  JHEP {\bf 0711}, 076 (2007)
  %doi:10.1088/1126-6708/2007/11/076
  [arXiv:0706.3903 [hep-th]].
  %%CITATION = doi:10.1088/1126-6708/2007/11/076;%%

%\cite{Buchbinder:2007at}
\bibitem{Buchbinder:2007at}
  E.~I.~Buchbinder, J.~Khoury and B.~A.~Ovrut,
  ``Non-Gaussianities in new ekpyrotic cosmology,''
  Phys.\ Rev.\ Lett.\  {\bf 100}, 171302 (2008)
  %doi:10.1103/PhysRevLett.100.171302
  [arXiv:0710.5172 [hep-th]].
  %%CITATION = doi:10.1103/PhysRevLett.100.171302;%%

%\cite{Lehners:2007wc}
\bibitem{Lehners:2007wc}
  J.~L.~Lehners and P.~J.~Steinhardt,
  ``Non-Gaussian density fluctuations from entropically generated curvature perturbations in Ekpyrotic models,''
  Phys.\ Rev.\ D {\bf 77}, 063533 (2008)
  Erratum: [Phys.\ Rev.\ D {\bf 79}, 129903 (2009)]
  %doi:10.1103/PhysRevD.79.129903, 10.1103/PhysRevD.77.063533
  [arXiv:0712.3779 [hep-th]].
  %%CITATION = doi:10.1103/PhysRevD.79.129903, 10.1103/PhysRevD.77.063533;%%

%\cite{Lehners:2008my}
\bibitem{Lehners:2008my}
  J.~L.~Lehners and P.~J.~Steinhardt,
  ``Intuitive understanding of non-gaussianity in ekpyrotic and cyclic models,''
  Phys.\ Rev.\ D {\bf 78}, 023506 (2008)
  Erratum: [Phys.\ Rev.\ D {\bf 79}, 129902 (2009)]
  %doi:10.1103/PhysRevD.78.023506, 10.1103/PhysRevD.79.129902
  [arXiv:0804.1293 [hep-th]].
  %%CITATION = doi:10.1103/PhysRevD.78.023506, 10.1103/PhysRevD.79.129902;%%

%\cite{Lehners:2009ja}
\bibitem{Lehners:2009ja}
  J.~L.~Lehners and S.~Renaux-Petel,
  ``Multifield Cosmological Perturbations at Third Order and the Ekpyrotic Trispectrum,''
  Phys.\ Rev.\ D {\bf 80}, 063503 (2009)
  %doi:10.1103/PhysRevD.80.063503
  [arXiv:0906.0530 [hep-th]].
  %%CITATION = doi:10.1103/PhysRevD.80.063503;%%

%\cite{Lehners:2008vx}
\bibitem{Lehners:2008vx}
  J.~L.~Lehners,
  ``Ekpyrotic and Cyclic Cosmology,''
  Phys.\ Rept.\  {\bf 465}, 223 (2008)
  %doi:10.1016/j.physrep.2008.06.001
  [arXiv:0806.1245 [astro-ph]].
  %%CITATION = doi:10.1016/j.physrep.2008.06.001;%%

%\cite{Lehners:2010fy}
\bibitem{Lehners:2010fy}
  J.~L.~Lehners,
  ``Ekpyrotic Non-Gaussianity: A Review,''
  Adv.\ Astron.\  {\bf 2010}, 903907 (2010)
  %doi:10.1155/2010/903907
  [arXiv:1001.3125 [hep-th]].
  %%CITATION = doi:10.1155/2010/903907;%%

%\cite{Fertig:2013kwa}
\bibitem{Fertig:2013kwa}
  A.~Fertig, J.~L.~Lehners and E.~Mallwitz,
  ``Ekpyrotic Perturbations With Small Non-Gaussian Corrections,''
  Phys.\ Rev.\ D {\bf 89}, no. 10, 103537 (2014)
  %doi:10.1103/PhysRevD.89.103537
  [arXiv:1310.8133 [hep-th]].
  %%CITATION = doi:10.1103/PhysRevD.89.103537;%%

%\cite{Ijjas:2014fja}
\bibitem{Ijjas:2014fja}
  A.~Ijjas, J.~L.~Lehners and P.~J.~Steinhardt,
  ``General mechanism for producing scale-invariant perturbations and small non-Gaussianity in ekpyrotic models,''
  Phys.\ Rev.\ D {\bf 89}, no. 12, 123520 (2014)
  %doi:10.1103/PhysRevD.89.123520
  [arXiv:1404.1265 [astro-ph.CO]].
  %%CITATION = doi:10.1103/PhysRevD.89.123520;%%

%\cite{Levy:2015awa}
\bibitem{Levy:2015awa}
  A.~M.~Levy, A.~Ijjas and P.~J.~Steinhardt,
  ``Scale-invariant perturbations in ekpyrotic cosmologies without fine-tuning of initial conditions,''
  Phys.\ Rev.\ D {\bf 92}, no. 6, 063524 (2015)
  %doi:10.1103/PhysRevD.92.063524
  [arXiv:1506.01011 [astro-ph.CO]].
  %%CITATION = doi:10.1103/PhysRevD.92.063524;%%

%\cite{Fertig:2015ola}
\bibitem{Fertig:2015ola}
  A.~Fertig and J.~L.~Lehners,
  ``The Non-Minimal Ekpyrotic Trispectrum,''
  JCAP {\bf 1601}, no. 01, 026 (2016)
  %doi:10.1088/1475-7516/2016/01/026
  [arXiv:1510.03439 [hep-th]].
  %%CITATION = doi:10.1088/1475-7516/2016/01/026;%%

%\cite{Fertig:2016czu}
\bibitem{Fertig:2016czu}
  A.~Fertig, J.~L.~Lehners, E.~Mallwitz and E.~Wilson-Ewing,
  ``Converting entropy to curvature perturbations after a cosmic bounce,''
  JCAP {\bf 1610}, no. 10, 005 (2016)
  %doi:10.1088/1475-7516/2016/10/005
  [arXiv:1607.05663 [hep-th]].
  %%CITATION = doi:10.1088/1475-7516/2016/10/005;%%

%\cite{Levy:2016xcl}
\bibitem{Levy:2016xcl}
  A.~M.~Levy,
  ``Fine-tuning challenges for the matter bounce scenario,''
  Phys.\ Rev.\ D {\bf 95}, no. 2, 023522 (2017)
  %doi:10.1103/PhysRevD.95.023522
  [arXiv:1611.08972 [gr-qc]].
  %%CITATION = doi:10.1103/PhysRevD.95.023522;%%

%\cite{Creminelli:2004yq}
\bibitem{Creminelli:2004yq}
  P.~Creminelli and M.~Zaldarriaga,
  ``Single field consistency relation for the 3-point function,''
  JCAP {\bf 0410}, 006 (2004)
  %doi:10.1088/1475-7516/2004/10/006
  [astro-ph/0407059].
  %%CITATION = doi:10.1088/1475-7516/2004/10/006;%%

%\cite{Cheung:2007sv}
\bibitem{Cheung:2007sv}
  C.~Cheung, A.~L.~Fitzpatrick, J.~Kaplan and L.~Senatore,
  ``On the consistency relation of the 3-point function in single field inflation,''
  JCAP {\bf 0802}, 021 (2008)
  %doi:10.1088/1475-7516/2008/02/021
  [arXiv:0709.0295 [hep-th]].
  %%CITATION = doi:10.1088/1475-7516/2008/02/021;%%
\end{thebibliography}
\end{document}